\documentclass[twocolumn]{aastex701}

\newcommand{\prot}{$P_{\rm rot}$}
\newcommand{\ktwo}{K2}
\newcommand{\kepler}{Kepler}
\newcommand{\gaia}{Gaia}
\newcommand{\teff}{$T_{\rm eff}$}
\newcommand{\tarsnrots}{\added{1,046,317}}

\usepackage{hyperref}
\usepackage{footmisc}
\usepackage{tikz}
\usepackage{graphicx}
\usepackage{sidecap}
\usepackage{rotating}
\usepackage{caption}
\usepackage{booktabs}
\usetikzlibrary{shapes.geometric, arrows, positioning}
\usepackage{longtable}
\usepackage{ulem}
\usepackage{amsmath}
\usepackage{enumitem}

\tikzstyle{startstop} = [rectangle, rounded corners, minimum width=2.5cm, minimum height=0.8cm, text centered, draw=black, fill=gray!10]
\tikzstyle{process} = [rectangle, minimum width=2.5cm, minimum height=0.8cm, text centered, draw=black, fill=white]
\tikzstyle{decision} = [diamond, aspect=2.5, minimum width=1cm, minimum height=0.5cm, text centered, draw=black, fill=white, inner sep=1pt]
\tikzstyle{arrow} = [thick,->,>=stealth]

\begin{document}

\title{The TESS All-Sky Rotation Survey: Periods for \added{\tarsnrots} Stars Within 500 pc}

\author[orcid=0000-0001-6037-2971]{Andrew W. Boyle}
\altaffiliation{NSF Graduate Research Fellow}
\affiliation{Department of Physics and Astronomy, The University of North Carolina at Chapel Hill, Chapel Hill, NC 27599, USA}
\email[show]{awboyle@unc.edu}  

\author[orcid=0000-0002-0514-5538]{Luke G. Bouma}
\altaffiliation{Carnegie Fellow}
\affiliation{Observatories of the Carnegie Institution for Science, Pasadena, CA 91101, USA}
\email{lbouma@carnegiescience.edu}

\author[orcid=0000-0003-3654-1602]{Andrew W. Mann}
\affiliation{Department of Physics and Astronomy, The University of North Carolina at Chapel Hill, Chapel Hill, NC 27599, USA}
\email{awmann@unc.edu}

\correspondingauthor{Andrew Boyle}

\begin{abstract}

Stellar rotation is a fundamental tracer of stellar magnetic evolution, age, and activity, with broad implications for Galactic archaeology and exoplanet characterization. The Transiting Exoplanet Survey Satellite (TESS) provides high-precision time-series photometry across the sky, enabling rotation measurements for an unprecedented number of stars. We present the TESS All-Sky Rotation Survey (TARS), an all-sky catalog of stellar variability periods for \tarsnrots\ stars with $T<16$ and distances within 500~pc. We estimate that \added{93\%} of these periods are rotation periods. This catalog increases the number of rotation period measurements for stars with $T < 16$ within 100~pc by a factor of \added{2.3} and within 500~pc by \added{4.0}. We also present a method to correct half-period \added{harmonics} in TESS data and show that it reliably recovers periods as long as 25 days from a single TESS sector. TARS represents the largest homogeneous catalog of stellar rotation periods to date, providing a foundation for studies of stellar ages, young associations, and Galactic structure. We make the light curves used in our analysis available as a HLSP through MAST\footnote{\url{https://archive.stsci.edu/hlsp/tars}}. Beyond the default TARS catalog, we provide code that allows users to generate rotation period catalogs with adjustable completeness and reliability thresholds. This code and all rotation period measurements are available through Zenodo\footnote{\href{https://doi.org/10.5281/zenodo.19917941}{10.5281/zenodo.19917941}}.

\end{abstract}

\section{Introduction} 

Stellar rotation is a key observable that traces the magnetic and structural evolution of low-mass stars. Young stars form with a wide range of rotation periods \citep[][]{rebullRotationLowmassStars2018} and spin up as they contract toward the zero-age main sequence. Stars with convective envelopes subsequently undergo magnetized-wind braking that steadily removes their rotational angular momentum \citep{schatzmanTheoryRoleMagnetic1962,weberAngularMomentumSolar1967, skumanichTimeScalesCA1972}. These processes link surface rotation to stellar age, magnetic activity, and internal structure, making rotation periods essential diagnostics across the main sequence.

In exoplanet science, stellar rotation presents both opportunities and challenges.  Since stellar rotation encodes the ages of main-sequence stars \citep{Barnes2007,Bouma2023}, it provides a path toward studying exoplanet evolution over time.  This is most relevant for evolutionary processes that alter planets over the first billion years \citep[e.g.,][]{OwenAtmospheric2019,Dai2024,Vach2024,Livingston2026}.  A key observational challenge, however, is that the transit and radial velocity methods are biased against active stars.  The community has developed creative approaches to this issue for both transits \citep{Jenkins:2010qy,Rizzuto2017,Hippke2019,Garcia2024} and radial velocities \citep{Cale2021,Klein2022,Barragan2022,Damasso2023}, enabling progress despite these challenges.

Over the past decade, wide-field photometric surveys have enabled rotation measurements for tens of thousands of stars. Kepler established that the rotation distribution of field K and M dwarfs is bimodal, both in the Kepler field \citep{mcquillanROTATIONPERIODS342014} and across the ecliptic \citep{reinholdStellarRotationPeriods2020}. Ground-based studies have measured rotation periods for nearby stars \citep{Binks2015,Jayasinghe2018}, open clusters \citep{Fritzewski2021, Dungee2022}, and M dwarfs \citep{Newton2016, Lu2022}. 

The Transiting Exoplanet Survey Satellite \citep[TESS;][]{Ricker2015} provides a unique opportunity to build on these previous studies. 
TESS provides high-quality photometry for nearly every star with $T \lesssim 16$ across the sky, enabling rotation measurements over a sample that no previous survey could approach. Although the 27-day observing window of a single sector limits sensitivity to periods beyond $\simeq12$\,days, TESS is still sensitive to the fast and moderate rotators that dominate young populations, and there has been some success extending this 12-day barrier \citep{Claytor_2024,Hattori2025}. Crucially, TESS’s all-sky footprint allows rotation to be mapped across Galactic environments rather than restricted to individual fields or clusters.

This all-sky capability has already begun to reshape our understanding of stellar populations \citep{curtisTESSRevealsThat2019,Kounkel2022,Wood2023}. For example, \citet{Boyle2025b} used TESS rotation periods and \textit{Gaia} astrometry to show that the Pleiades is part of a much larger complex of associated stars. Such studies illustrate the crucial role of all-sky rotational catalogs. 

\begin{figure*}
    \centering
    \includegraphics[width=1\linewidth]{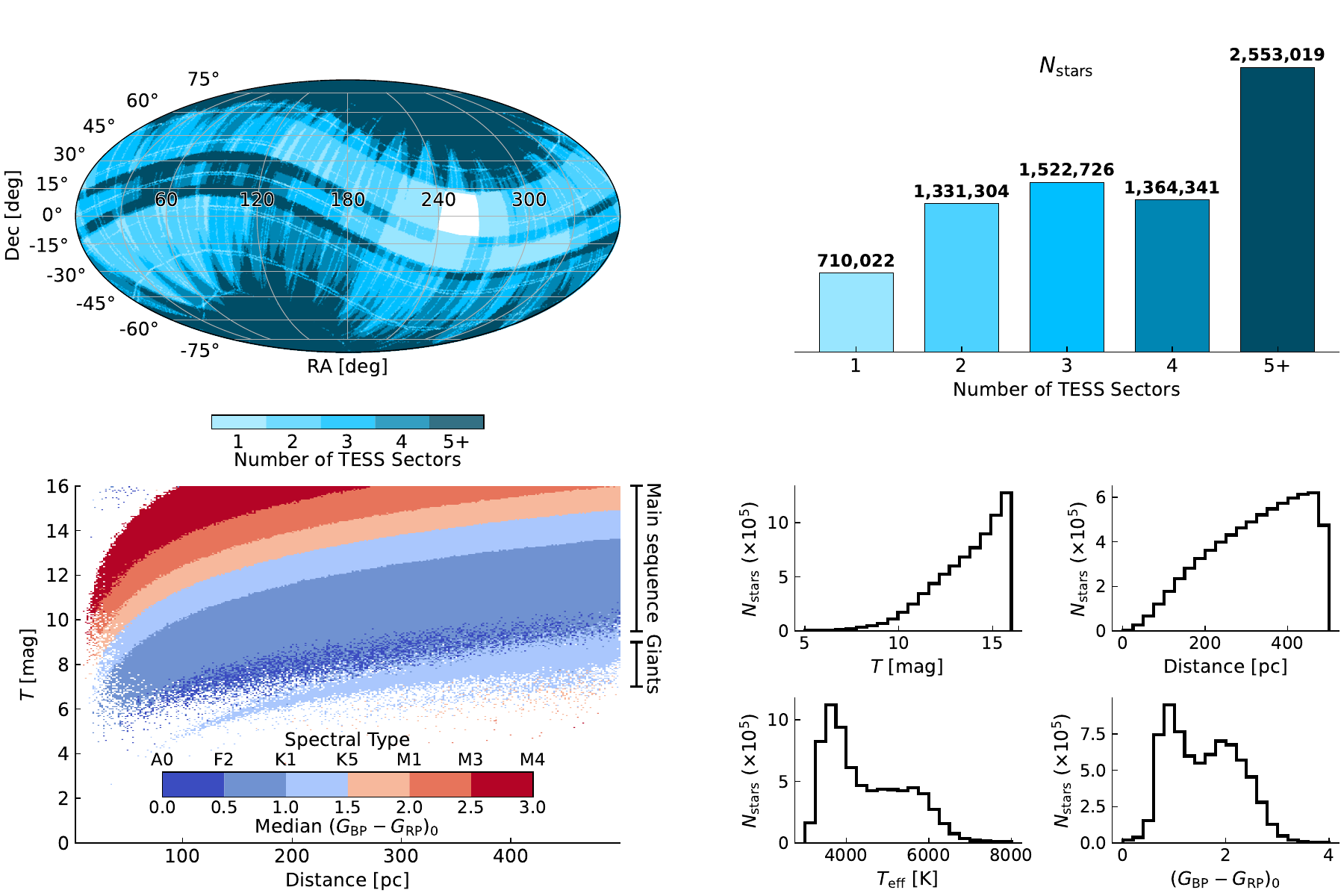}
    \caption{ {\bf Selection function.}
    \textit{Top left:} Sky coverage. We used TESS images from 2018~July--2025~September; the white gap is where TESS did not observe in Sectors 1--96.
    \textit{Top right:} Histogram of number of TESS observations per star; numbers above each bin give the star count. The median star has four sectors of TESS data. 
    \textit{Lower left:} The TESS magnitude $T$ as a function of the distance to each star in our survey, colored by the median de-reddened \gaia\ DR3 $G_{\rm BP} - G_{\rm RP}$ color in each bin. Spectral types are approximated from \citet{2013ApJS..208....9P}.
    \textit{Lower right:} Histograms across our sample of magnitude, distance, temperature, and color. 
    }
    \label{fig:sky_plot}
\end{figure*}

To date, no existing study has produced a flux- and distance-limited catalog of TESS rotation periods. Previous TESS rotation efforts have focused on specific clusters \citep{THYMEVII,Douglas2024,Stafford2026} and known planet hosts \citep{CantoMartins2020, Stelzer2022}. Broader studies by \citet{2023ApJS..268....4F} and \citet{2024AJ....167..189C} have used two-minute data from the TESS prime mission to identify tens of thousands of new variable stars.  However, most stars within 500\,pc of the Sun have not been observed by TESS at two-minute cadence due to the heterogeneous nature of the two-minute data product \citep{Fausnaugh2021}.   In contrast, the TESS full-frame images (FFIs) offer an opportunity to homogeneously survey the local neighborhood for stellar variability. 

In this paper, we present the TESS All-Sky Rotation Survey (TARS): an all-sky catalog of rotation periods for stars within 500 pc with $T < 16$. We emphasize transparent selection functions and provide multiple catalog layers that allow users to trade completeness against reliability depending on their needs. We aim to provide a community resource that not only reports TESS rotation periods but also maps reliability and completeness across the sky.  This enables all-sky studies of stellar ages, young associations, and Galactic structure, and supports improved exoplanet characterization through reliable rotation constraints.

\section{Target Selection} \label{sec:target_selection}

We aim to measure reliable stellar rotation periods for as large and diverse a sample as possible. 
Our selection function is therefore guided by three considerations: (1) maximizing the number of stars with measurable rotation periods, (2) ensuring that TESS has sufficient photometric sensitivity to recover those periods, and (3) achieving broad coverage of nearby, known stellar associations to use as validation benchmarks.
Given these considerations, we chose to focus on stars observed by TESS with $T < 16$ and $d < 500$~pc.
We constructed the resulting target list as follows.

\begin{enumerate}[leftmargin=12pt,topsep=0pt,itemsep=0ex,partopsep=1ex,parsep=1ex]
    \item We began with stars in TIC v8.2 \citep{2021arXiv210804778P} satisfying $T < 16$ and \gaia\ DR2 $\varpi > 2$~mas. This initial cut yielded 7,914,103 stars.
    
    \item We removed sources flagged as duplicates in TIC v8.2, eliminating 127,205 stars.
    
    \item We excluded stars lacking an associated \gaia\ DR2 source identifier in the TIC, removing 986 stars.
    
    \item For the remaining targets, we identified corresponding \gaia\ DR3 sources using the \gaia\ DR3 \texttt{dr2\_dr3\_neighbourhood} table. We required matches within $2\arcsec$ and a magnitude difference $\Delta G < 0.5$~mag. When multiple candidate matches satisfied these criteria, we adopted the DR3 source with the smallest angular separation. The cross-matched sample contained 7,710,873 stars.
    
    \item We removed stars with missing \gaia\ DR3 $G$, $BP$, or $RP$ photometry, yielding 7,655,488 stars.
    
    \item We then used \texttt{tess-point} \citep{2020ascl.soft03001B} to determine whether each star fell on TESS silicon. This process yielded 7,481,412 stars observed by TESS at least once between Sectors~1--96 (2018~July--2025~September).
\end{enumerate}

We adopted this final set of 7,481,412 stars as our target sample. We show general information about these targets in Figure~\ref{fig:sky_plot}.  Our selection function includes 97\% of AFGK dwarfs out to 500\,pc, and a similar fraction of partially-convective M dwarfs out to $\simeq$250\,pc. Most targets have multiple sectors of data, corresponding to a total of 39,061,674 star--sector observations.

For this sample, we corrected the photometry for interstellar extinction and computed effective temperatures (\teff) following \citet{Boyle2025b}. In brief, this procedure involved combining the three-dimensional dust maps from \citet{Vergely2022} with the \gaia\ DR3 extinction coefficients from \citet{Fitzpatrick2019}. We then estimated effective temperatures (\teff) using an updated version of the \gaia\ DR2 color--\teff{} relation from \citet{Curtis2020}, recalibrated for \gaia\ DR3 photometry.

\section{Generating Light Curves} \label{sec:lc_generation}

This work is focused on measuring rotation periods for stars brighter than $T < 16$ within 500\,pc.  Given the available TESS data products at the time of our analysis, the best match for our needs was the full-frame images.  While a wide variety of data processing efforts have yielded TESS light curves \citep{jenkinsTESSScienceProcessing2016,2020MNRAS.498.5972N,Bouma2019,2020RNAAS...4..204H,Caldwell2020,Han2023,Hartman2025}, the selection functions of these analyses and their availability through MAST did not match our requirements.

\begin{figure*}
    \centering
    \includegraphics[width=0.99\linewidth]{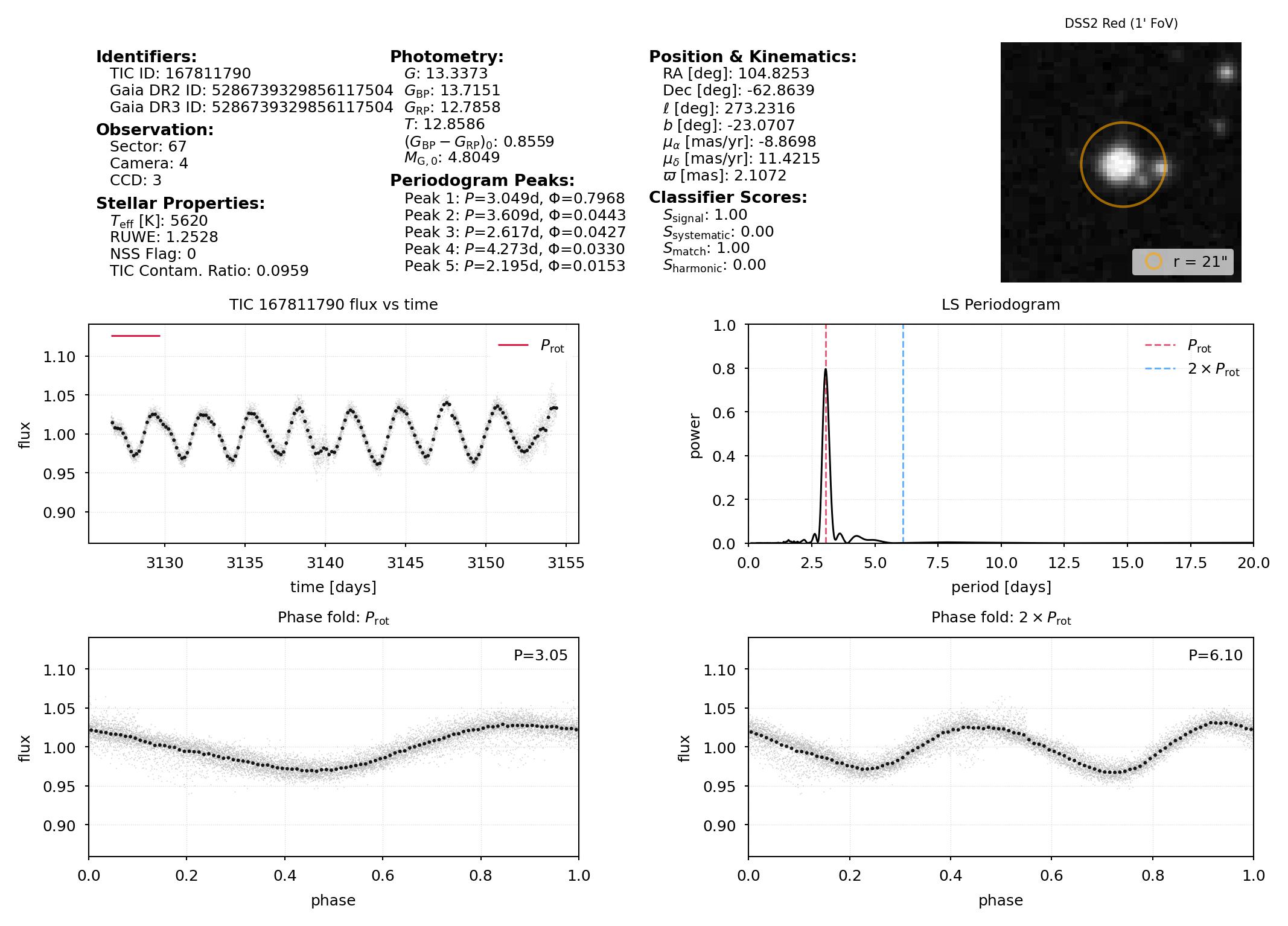}
    \caption{{\bf Example vetting plot.}  Probabilities from our systematics and \added{harmonic} classifiers are listed.  One such vetting plot is available for each of the 39,044,569 light curves through the TARS High-Level Science Product at MAST.  An interactive vetting plot explorer is also {\bf \href{https://lgbouma.com/tars_viz/}{available online}}.
    }
    \label{fig:first_stage_lcs}
\end{figure*}

The \texttt{unpopular} package \citep{hattoriUnpopularPackageDatadriven2021} enables light curve extraction from FFIs and has a proven track record in studies of stellar rotation \citep[e.g.,][]{Boyle2025b, Hattori2025, Stafford2026}. To remove TESS systematics, \texttt{unpopular} uses a causal pixel model \citep{Wang2016}, which assumes that signals common to many pixels within an image are instrumental, while signals unique to individual pixels are astrophysical. A systematics model is constructed from these shared signals and removed from each target light curve. The package is open-source\footnote{\url{https://github.com/soichiro-hattori/unpopular}} and straightforward to deploy, making it well suited for large-scale surveys. We used the \texttt{unpopular} version corresponding to Git commit \texttt{5653498}.

We used \texttt{unpopular} following the same setup as in \citet{Boyle2025a}. Briefly, for each star and sector, we used \texttt{TESSCut} to extract a $50\times50$ pixel cutout from the calibrated FFIs centered on the target coordinates. Prior to systematics correction, we subtracted a background estimated from the median counts of the 300 faintest pixels in each image. We then constructed a systematics model using light curves from 100 pixels of similar brightness, excluding all pixels within 5 pixels of the target. \texttt{unpopular} prevents overfitting by using a regularization parameter ($\lambda$) that constrains the model flexibility, which we set to 0.1.  We then divided each light curve into 100 temporal segments, trained the systematics model on each of the other segments, and applied the resulting best-fit coefficients to the holdout segment. Subtracting this model yielded the final corrected light curve for each star.

We applied this procedure to all 39,061,674 star–sector combinations in our survey, yielding one light curve for each star in each sector in which it was observed. \added{\texttt{TESSCut} returned empty flux arrays for 17,105 light curves. These empty light curves were removed, yielding \added{39,044,569} light curves for analysis.}

\section{Methods} \label{sec:methods}

\subsection{Measuring Periodic Signals} \label{subsec:measuring_prot}

For each of our \added{39,044,569} light curves, we used a Lomb-Scargle (LS) periodogram \citep{lombLeastsquaresFrequencyAnalysis1976, scargleStudiesAstronomicalTime1982} as implemented in \texttt{astropy} (\!\!\citealt{astropy:2013, astropy:2018, astropy:2022}) to search for periodicity. The details of our implementation and all parameters we recorded are given in Appendix~\ref{sec:measurements}. These measurements can be accessed and downloaded from Zenodo\footnote{\href{https://doi.org/10.5281/zenodo.19917941}{10.5281/zenodo.19917941}}. Figure~\ref{fig:first_stage_lcs} shows an example light curve and periodogram. Similar plots for all rotation measurements in the TARS survey are available at MAST\footnote{\url{https://archive.stsci.edu/hlsp/tars}}.

\subsection{Vetting Periodic Signals} \label{subsec:validating_prot}

The best-fit frequency in a Lomb-Scargle periodogram corresponds to the strongest sinusoidal signal in the light curve, but this signal does not necessarily correspond to stellar variability. In particular, instrumental systematics---including momentum dumps and scattered light---can produce prominent periodic signals. In addition, \kepler\ revealed numerous FGK stars with rotation periods of 20-30 days \citep{mcquillanROTATIONPERIODS342014}. While the amplitudes of such signals might be detected by TESS, the periods themselves are usually resolved as the half-period harmonic. 

We addressed the issues of systematics classification and \added{harmonic} identification using two sequential random forest classifiers. Random forest classifiers have previously been shown to perform well in identifying stellar rotation periods \citep{santosSurfaceRotationPhotometric2019, santosSurfaceRotationPhotometric2021b, 2024A&A...689A.229B}, and behave well for large datasets.  We designed our first classifier (hereafter: the systematics classifier) to separate instrumental systematics from \added{persistent} signals.   We trained the second classifier (hereafter: the \added{harmonic} classifier) to separate half-period \added{harmonics} from accurately-measured rotation signals.  We describe each in detail below.

\subsubsection{The Systematics Classifier} \label{subsubsec:systematics}

\begin{figure*}
    \centering
    \includegraphics[width=1\textwidth]{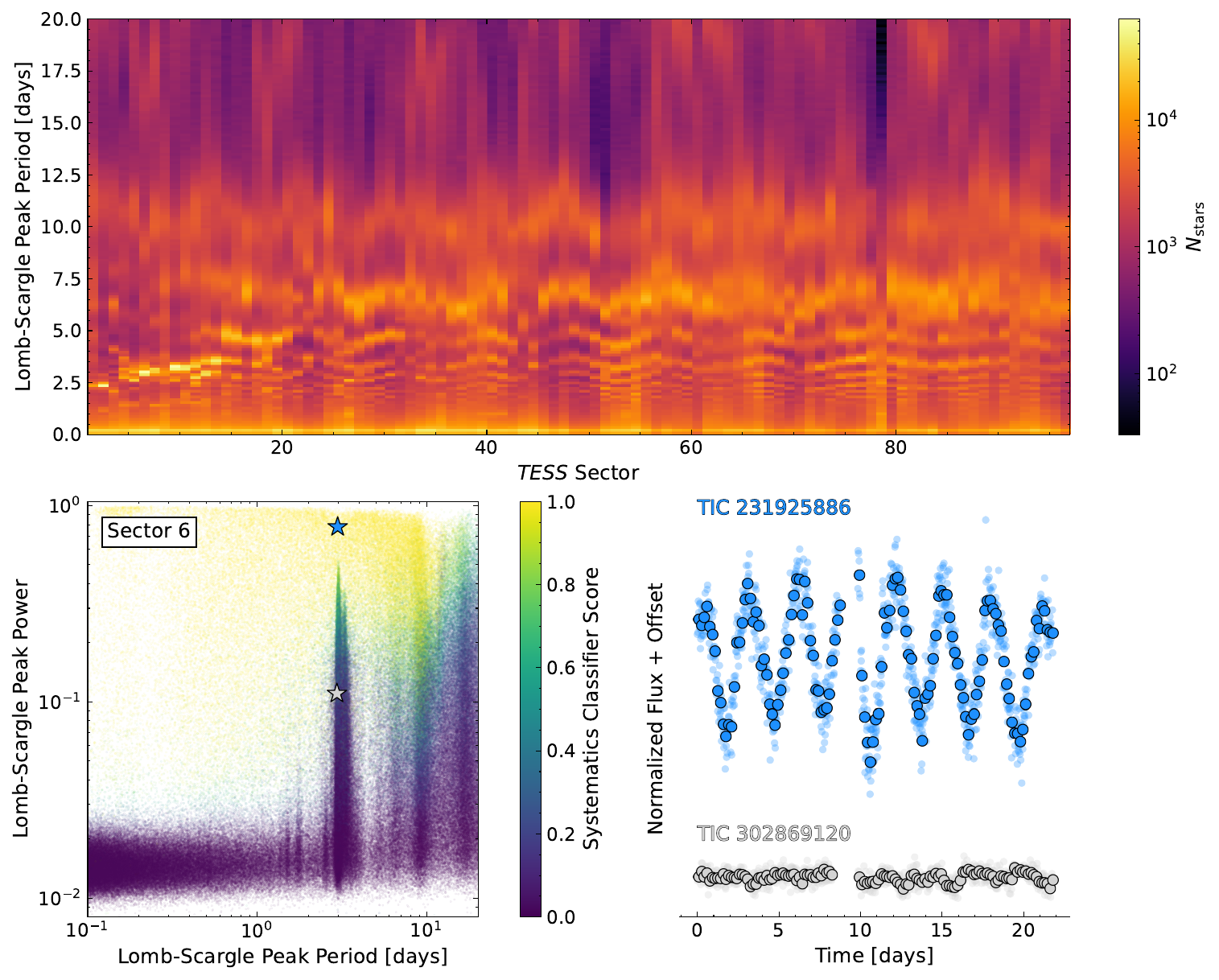}
    \caption{\textbf{Systematics masquerade as period detections.} \textit{Top:} The period of the highest peak in the Lomb-Scargle periodogram across all TESS sectors. The sinusoidal structures across sectors are harmonics of the TESS data gap. 
    \textit{Bottom left:} Our systematics classifier identifies and separates systematics from other rotation signals. The systematics classifier is trained so rotation measurements with a high probability are more likely to be real signals and less likely to be a systematic. Sector~6 experienced momentum dumps every 3.125 days. The blue and gray stars show the Lomb-Scargle period and power of the two light curves in the bottom right panel.
    \textit{Bottom right:} TIC 231925886 and TIC 302869120 both have Lomb-Scargle peak periods of $\sim2.9$ days in Sector 6, but TIC 231925886 has a systematics classifier score of 1.0 (high probability of being a rotation signal)  while TIC 302869120 has a systematics classifier score of 0.0 (high probability of being a systematic). 
    }
    \label{fig:systematics}
\end{figure*}

The most common systematics in TESS light curves are scattered light from the Earth and Moon, and spacecraft momentum dumps.
The average TESS light curve therefore has excess power at harmonics of the spacecraft orbital period, and at harmonics of the time interval between momentum dumps (Figure~\ref{fig:systematics}).   Prior work has taken advantage of this period-clustering to remove non-astrophysical systematics \citep[][]{2023ApJS..268....4F}. 
Signal amplitudes offer an independent dimension to separate signal from noise.
Our approach, beyond recording Lomb-Scargle peak periods and the amplitudes of the associated best-fit Fourier models, is to compute a few dozen light curve and periodogram features (see Appendix~\ref{sec:measurements}), and to use them as inputs for our systematics classifier.

\begin{figure*}[ht]
    \centering
    \includegraphics[width=0.98\linewidth]{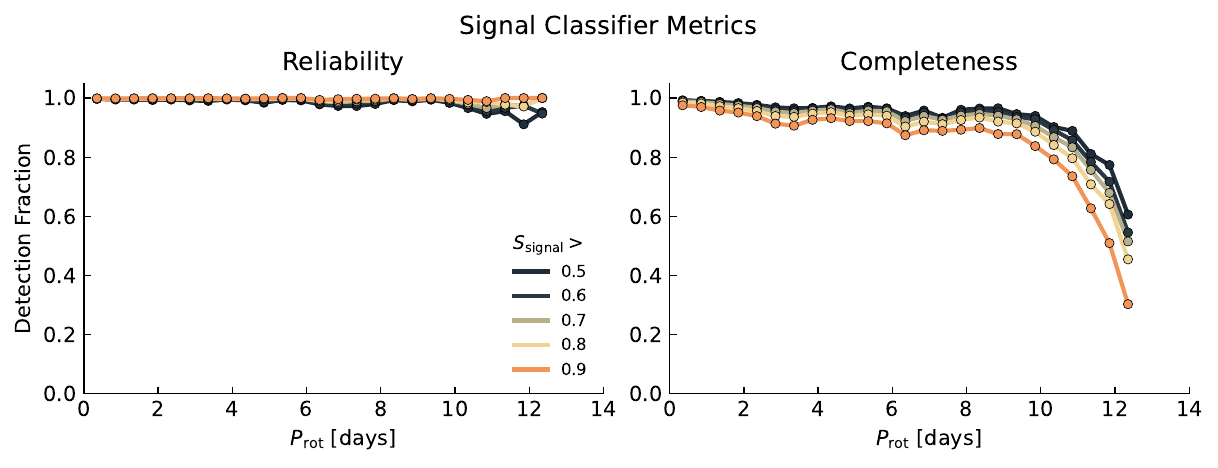}
    \includegraphics[width=0.98\linewidth]{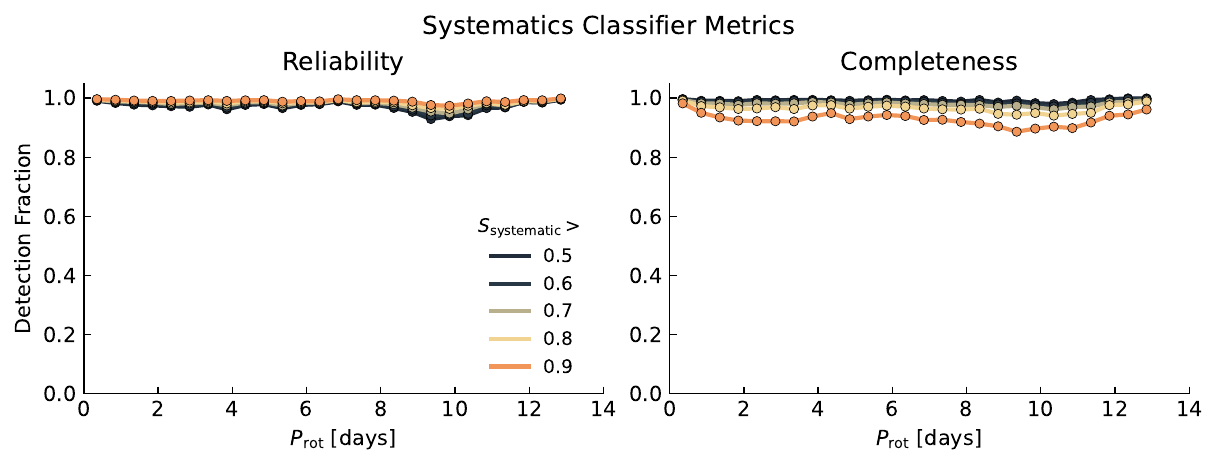}
    \caption{{\bf Systematics classifier performance vs.~rotation period.}
    \textit{Top left:} Reliability, defined as the fraction of random forest-selected signals that yield rotation periods consistent across $>$$5$ sectors, shown for several random forest probability thresholds. The classifier returns reliable detections across the full period range. 
    \textit{Top right:} Completeness, defined as the fraction of all signals in the class that exceed the random forest probability threshold cuts. Completeness remains high for $P_{\rm rot}\lesssim 10$~days, but decreases at longer periods. These panels suggest that the systematics classifier is reliable at longer periods, however the yield of such measurements declines beyond $\sim$10--13~days.
    \textit{Bottom left}: The reliability of the negative class as a function of $S_{\rm signal}$.
    \textit{Bottom right:} The completeness of the negative class as a function of \added{$S_{\rm systematic}$}.
    }
    \label{fig:first_thresholds}
\end{figure*}

We trained the systematics classifier using two labeled samples: a set of period measurements that we deemed reliable (the positive class) and a set that were dominated by systematics (the negative class).
We opted to define our positive and negative classes using internal consistency across multiple TESS sectors. For the positive class, we selected stars observed in five or more TESS sectors for which the measured period agreed across all sectors within $3\sigma$, with uncertainties from Equation~1 of \citet{Boyle2025a}.  For the negative class, we selected stars also observed in five or more sectors but for which the inferred periods were inconsistent between every sector. These measurements were therefore more likely to be noise than true \added{stellar variability}. This procedure yielded 508,455 measurements in the positive class (persistent signals) and 1,490,169 measurements in the negative class (inconsistent signals). To balance the training data, we randomly drew from the negative class to match the size of the positive class. \added{We then split our data and used 80\% to train the classifier, reserving 20\% for validation. This classifier is designed to identify persistent signals, but we note that persistence alone does not guarantee that a signal is astrophysical.}

We implemented a random forest classifier using \texttt{sklearn}'s \texttt{RandomForestClassifier} with 100 estimators, reserving 20\% of the data for validation \citep{scikit-learn}.  We show classifier performance and feature importance diagnostics in Appendix~\ref{sec:rf_val_plots}. The results were insensitive to the number of estimators and other hyperparameter variations, so we adopted the default settings. To guide feature selection, we introduced a control variable constructed by randomly permuting the measured Lomb-Scargle peak-power values among stars, and computed its feature importance.  All adopted features had higher importance than this randomized control.

\added{For the rest of this paper and in our plots, we refer to measurements with a high chance from this classifier of being a \added{persistent} signal as having a high $S_{\rm signal}$. Measurements that are more likely to be a systematic have a high $S_{\rm systematic}$. Since this is a binary classification problem, these quantities are related by $S_{\rm systematic} = 1 - S_{\rm signal}$. We note that these classifier scores are not true probabilities but that a high score of being a signal still means that the measurement is more likely to be a real signal.}

When evaluated on the validation set, the classifier correctly identifies 99\% of the negative class and 98\% of the positive class. The corresponding false positive and false negative rates are 0.9\% and 2\%, respectively. As an additional performance metric, we compute the area under the receiver operating characteristic (ROC) curve (AUC), where an AUC of 0.5 indicates no predictive power and an AUC of 1 corresponds to a perfect classifier. Our classifier achieves an AUC of 0.998, indicating excellent performance. The overall accuracy (the proportion of all classifications that are correct), precision (the proportion of \added{signals} classified as reliable that are actually reliable), and recall (the proportion of reliable \added{signals} classified as reliable) scores were 98.54\%, 99.11\%, and 97.95\%, respectively. As shown in Appendix~\ref{sec:rf_val_plots}, the classifier relies most strongly on the signal-to-noise ratios of the first five periodogram peaks, the amplitude of the best-fitting sinusoid, and the width of the highest periodogram peak.

The effectiveness of the classifier is visually apparent in the bottom-left panel of Figure~\ref{fig:systematics}. Using Sector~6 as an illustrative example, the low-probability points coincide with systematic-driven overdensities, where the vertical streak at $\sim3$ days is from momentum dumps.  We highlight this with two specific examples in the bottom-right panel of Figure~\ref{fig:systematics}. The top light curve, with a systematics classifier score of 1.0, exhibits clear, coherent rotational modulation that is readily identifiable by eye. The lower example, with a low \added{score} (0.0), shows low-amplitude variability that is ambiguous and difficult to confidently attribute to stellar rotation. Visual inspection of similar systems suggests the classifier is effectively separating out rotation from systematics.

\textbf{Defining the Signal Detection Threshold}: While we provide the means for anyone to select a rotation catalog optimized for their science (see Section~\ref{sec:conclusion}), we need to adopt a threshold for the standard data product. We also need to quantify the role a shifting detection threshold has on the sample. To this end, we employ a mix of (1) reliability and completeness diagnostics, and (2) manual vetting.

Following \citet{Boyle2025a}, we define reliability as the fraction of stars the systematics classifier correctly identifies as being in each class and completeness as the fraction of signals in each class that are successfully recovered. Figure~\ref{fig:first_thresholds} illustrates the tradeoff between reliability and completeness for the systematics classifier. Reliability remains high even down to a \added{$S_{\rm signal}>0.5$}, with a slight drop-off at the longest periods for thresholds $<$0.7. Completeness remains high for periods $\lesssim10$\,days, but declines at longer periods due to the duration of TESS single-sector observations. 

We manually vetted a randomly selected sample of $\sim4{,}000$ measurements. For each, we visually inspected the TESS light curve, the light curve phase-folded at the period corresponding to the highest Lomb-Scargle peak, and the periodogram itself. We found that periods associated with a systematics classifier threshold $\gtrsim 0.8$ were generally convincing, while measurements with \added{$0.5 < S_{\rm signal} < 0.8$} were often ambiguous and difficult to assess with confidence. Periods with \added{$S_{\rm signal} < 0.5$} were rarely judged to be reliable.

Based on the combined evidence from the reliability and completeness analysis (Figure~\ref{fig:first_thresholds}) and the manual vetting, we required that a measurement have a systematics classifier \added{score of $S_{\rm signal} > 0.8$} to pass this stage of vetting. Applying this cut to our hold out validation sample, we find that 99.69\% of the measurements that pass this threshold are reliable and we still capture 96.29\% of reliable detections. 

\subsubsection{The \added{Harmonic} Classifier}\label{subsubsec:alias_rf}

\begin{figure*}
    \centering
    \includegraphics[width=1\linewidth]{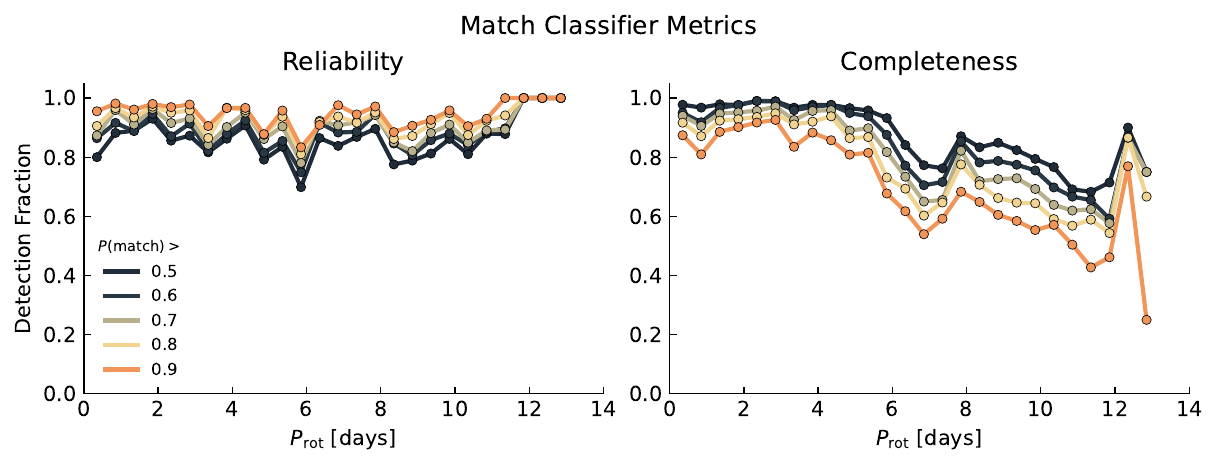}
    \includegraphics[width=1\linewidth]{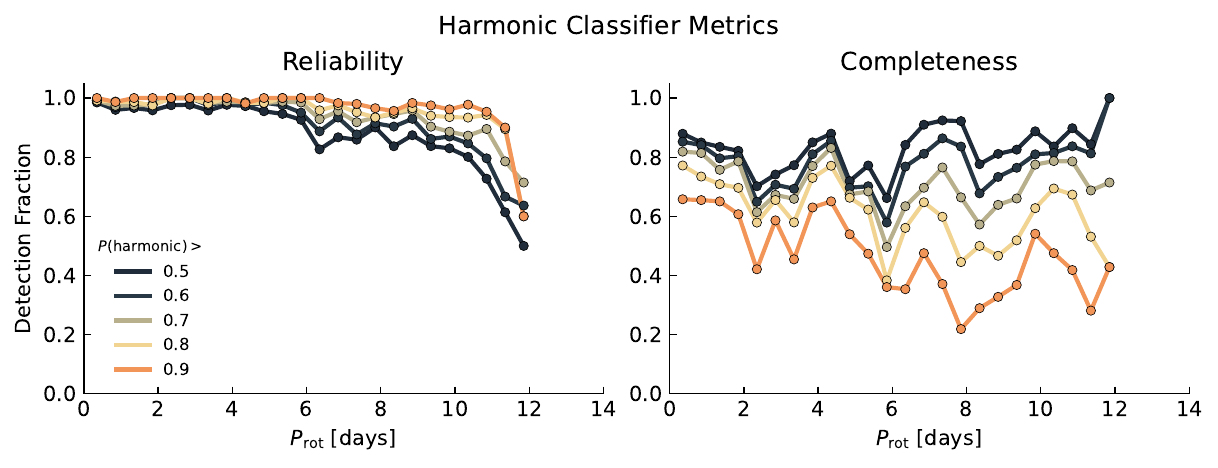}
    \caption{
    \textbf{Half-period harmonics can be identified at the cost of completeness.}
    \textit{Top left:} Reliability, defined here as the fraction of retained measurements corresponding to the true rotation period rather than a harmonic, remains uniformly high across the period range probed by TESS. Our ground truth periods are taken from \citet{mcquillanROTATIONPERIODS342014}.
    \textit{Top right:} This gain in reliability comes at the expense of completeness, which decreases steadily for more stringent cuts.
    \textit{Bottom left:} The random forest classifier likewise identifies half-period harmonics robustly across period space, with high reliability indicating that most measurements flagged as harmonics are indeed harmonics.
    \textit{Bottom right:} As for the true-period selection, imposing a high probability threshold for harmonic identification reduces completeness, removing a large fraction of potential harmonic detections.
    }
    \label{fig:rf_alias_thresholds}
\end{figure*}

While the systematics classifier is effective at separating instrumental systematics from \added{likely} astrophysical variability, it does not identify period aliasing incurred by the TESS window function. To address this issue, we trained a second random forest classifier to identify half-period harmonics. \added{Harmonics exist in TESS data at integer multiples and submultiples of the true period, but half-period harmonics (i.e. when TESS recovers $P_{\rm rot}/2$ instead of $P_{\rm rot}$) are by far the most common \citep{Boyle2025a}. This means it is hard to collect enough training data to reliably identify other harmonics so we limit our classifier to identifying half-period harmonics of the true period}.

Since this is a binary classification problem with matches defined as the positive class and \added{harmonics} defined as the negative class, a high \added{score} from the \added{harmonic} classifier \added{means the measured period is more likely to be a match}, whereas a low \added{score} indicates that the measured period is \added{more likely to be} a \added{harmonic}. \added{We again denote these scores as $S_{\rm match}$ and $S_{\rm harmonic}$.}

For the \added{harmonic} classifier, we defined positive and negative training classes using the \kepler\ rotation period catalog of \citet{mcquillanROTATIONPERIODS342014} (hereafter M14). \kepler's four-year observing baseline and 95\,cm aperture yielded rotation periods that are well-suited to identifying half-period \added{harmonics}.
M14 reported 34,030 rotation periods spanning 0.2--70\,days out of a parent sample of 133,030 stars. We generated TESS light curves for this sample as in Section~\ref{sec:lc_generation}. Starting from the 34,030 stars with detected periods in M14, we determined which targets were observed by TESS using \texttt{tess-point}, and excluded stars that were not observed, lacked valid \gaia\ DR3 photometry, or had $T>16$. This yielded 312,860 TESS light curves corresponding to 32,684 unique stars. We then computed Lomb--Scargle periodograms and recorded the same set of parameters described in Appendix~\ref{sec:measurements}.

For each sector, we compared our measured period to the corresponding M14 period. We classified a measurement as a match if the periods agreed within $3\sigma$, as defined by Equation~1 of \citet{Boyle2025a}, and as a \added{harmonic} if the measured period agreed within $3\sigma$ with half of the M14 period. To mirror our full vetting pipeline, we then applied the systematics classifier described in Section~\ref{subsubsec:systematics} to this sample. A total of 47,708 measurements satisfied our quality cut of \added{$S_{\rm signal}>0.8$}, of which 34,513 were matches and 13,195 were \added{harmonics}. To balance the training data for the \added{harmonic} classifier, we then randomly selected 13,195 measurements from the matched sample. \added{We found that without down-sampling, this classifier tended to misclassify harmonics as matches at twice the rate as when we use a down-sampled data set. Given that the goal of this classifier is to identify accurately identify harmonics, we have chosen to use the down-sampled training set in our classifier.}

We again reserved 20\% of the data as a hold-out validation set and used K-Fold cross-validation to train the random forest model. We used \texttt{sklearn}'s \texttt{RandomForestClassifier} with \texttt{n\_estimators = 200}, \texttt{max\_features = 0.5}, and \texttt{criterion = entropy}. We left the rest of the values at their default \texttt{sklearn} values. We list the adopted features in Appendix~\ref{sec:measurements}, and their importances in Appendix~\ref{sec:rf_val_plots}.

Our \added{harmonic} classifier distinguishes true rotation periods from \added{harmonics} with high fidelity. It correctly identifies true rotation signals in 89\% of cases and correctly flags \added{harmonics} in 85\% of cases. \added{Harmonics} are incorrectly classified as true signals 11\% of the time (the false-positive rate), and true rotation periods are misclassified as \added{harmonics} in 15\% of cases (the false-negative rate). The discriminative performance of the classifier is strong, with an AUC of 0.951. The resulting classification achieves an accuracy of 87.3\%, with a precision of 85.9\% and a recall of 89.3\%. 
Among the input features, the difference in BIC between a two-term Lomb--Scargle model evaluated at the period of the highest periodogram peak and at twice that period provided the greatest discriminatory power. Other important features included the ratio of the first and second detected periods, the Lomb--Scargle power evaluated at twice the best period, and the period corresponding to the second-highest Lomb--Scargle peak. 

\added{
We take the additional step of using Bayes' theorem to convert the \added{harmonic} classifier scores into Bayesian probabilities. This lets us estimate the probability that the detected period is a true match or half-period \added{harmonic} and gives us each measurement's false positive probability, which we will use to calculate our adopted periods in the next section (Section~\ref{subsec:period_logic}). Specifically, for a measurement with detected period $P_{\rm rot}$ and \added{harmonic}-classifier score $S$, we estimate $P(\mathrm{match}\mid P_{\rm rot}, S_{\rm match})$ (the probability that the measurement is a match given its period and match classifier score) and $P(\mathrm{harmonic}\mid P_{\rm rot}, S_{\rm harmonic})$ (the probability that the measurement is a harmonic given its period and harmonic classifier score) from the holdout validation sample. For the rest of this paper, we refer to the Bayesian probability that a measurement is a match as $P(\rm match)$ and the probability that the measurement is an harmonic as $P(\rm harmonic)$.
}

{\bf Defining the Harmonic Detection Threshold:}~Figure~\ref{fig:rf_alias_thresholds} illustrates how reliability and completeness vary as a function of $P({\rm harmonic})$ for the harmonic classifier. For all tested thresholds, the minimum reliability exceeds 80\% at all periods. Adopting a stringent threshold (e.g., \added{$P({\rm match}) > 0.9$}) ensures that more than 94\% of the selected periods are not \added{harmonics}, independent of rotation period. The corresponding completeness, shown in the right panel, highlights the cost of such a cut: requiring a threshold $>0.9$ would remove up to half of the potential matches from the sample. These trends are summarized in Table~\ref{tab:rf_performance} and provide a quantitative guide to those who wish to adjust these settings (see Section~\ref{sec:conclusion}).

\begin{table}[!t]
\centering
\caption{Reliability and completeness for match and harmonic classification.}
\label{tab:rf_performance}
\vspace{0.3cm}
\resizebox{\columnwidth}{!}{%
\begin{tabular}{ccccc}
\hline
 & \multicolumn{2}{c}{Match} & \multicolumn{2}{c}{Harmonic} \\
Threshold & Reliability & Completeness & Reliability & Completeness \\
\hline
0.5 & 0.84 & 0.88 & 0.88 & 0.83 \\
0.6 & 0.87 & 0.85 & 0.91 & 0.77 \\
0.7 & 0.89 & 0.81 & 0.94 & 0.71 \\
0.8 & 0.92 & 0.77 & 0.97 & 0.61 \\
0.9 & 0.94 & 0.71 & 0.98 & 0.46 \\
\hline
\end{tabular}}
\vspace{0.5cm}
\end{table}

{\bf Adopted Detection Thresholds:}~Based on Figures~\ref{fig:first_thresholds} and~\ref{fig:rf_alias_thresholds}, we adopted a probability threshold of 0.8 for both matches and \added{harmonics}.  While this removes \added{up to 30\%} of rotation period detections, it ensures that $>95\%$ of our reported periods are the true rotation signal and not a \added{harmonic}.

\subsection{Period Adoption Logic}\label{subsec:period_logic}

The default TARS periods are intended to balance reliability and completeness. A user might be interested in a catalog that is more reliable but less complete, or vice versa. Our Zenodo\footnote{\href{https://doi.org/10.5281/zenodo.19917941}{10.5281/zenodo.19917941}} contains code and suggestions for how to take our catalog and make different period selections based on science needs.  We built our standard rotation period catalog as follows:

\begin{enumerate}[leftmargin=12pt,topsep=0pt,itemsep=0ex,partopsep=1ex,parsep=1ex]
    \item \textbf{Sector-level periods.}
    We began with the highest-power Lomb--Scargle peak (see Section~\ref{subsec:measuring_prot}) from the light curve for each sector of every target.

    \item \textbf{Period metrics.}
    For each sector of every target, we computed the metrics listed in Appendix~\ref{sec:measurements}.

    \item \textbf{Apply systematics classifier.}
    We removed any sector-level periods with $S_{\rm signal} < 0.8$ (Section~\ref{subsubsec:systematics}).

    \item \textbf{Flag half-period \added{harmonics}.}
    We flagged all targets with a \added{$P(\rm harmonic) > 0.8$} as likely \added{harmonics} and targets with a $P(\rm match) > 0.8$ as likely matches.

    \added{
    \item \textbf{Count \added{harmonic} evidence.}
    For each \added{harmonic}-flagged sector with detected period $P$, we counted the number of \added{harmonic}-flagged sectors that agreed with $P$ and match-flagged sectors that agreed with $2P$ within 15\%.
    
    \item \textbf{Count match evidence.}
    For the same candidate period $P$, we counted the number of match-flagged sectors that agreed with $P$ and \added{harmonic}-flagged sectors that agreed with $P/2$ within 15\%.


    \item \textbf{Apply doubling.}
    We compared the \added{harmonic} evidence against the match evidence after subtracting the number of sectors expected from false positives on each side, as estimated from the Bayesian probabilities calculated in Section~\ref{subsubsec:alias_rf}. If there were more \added{harmonic}-flagged sectors than would be expected by chance and the \added{harmonic} evidence exceeded the match evidence, we doubled the measured periods of the \added{harmonic}-flagged sectors in the corresponding period cluster.
    }
    

    \item \textbf{Assign uncertainties.} \label{item:uncertainties}
     We assigned each period an uncertainty using Equation~1 of \cite{Boyle2025a}.

    \item \textbf{Consistency check against the doubled-period ensemble.}
    We computed a weighted mean of the doubled periods (if any exist). Sector measurements that were \emph{not} doubled were retained as-is, and those that agreed with the doubled-period weighted mean to within 15\% were treated as consistent with the corrected period in subsequent averaging.

    \item \textbf{Outlier rejection.}
    We applied $3\sigma$-clipping (with the median absolute deviation as the scatter estimator) to the set of periods to remove discrepant sector-level measurements.

    \item \textbf{Compute the adopted period.}
    The adopted rotation period was then the weighted mean of the clipped periods. Our uncertainty was the standard error on the weighted mean.
\end{enumerate}

This procedure ensures that exactly one final rotation period was adopted for each star, while explicitly accounting for \added{harmonics}, inter-sector consistency, and measurement uncertainty. \added{The uncertainties we provide are statistical uncertainties; we do not consider systematic uncertainties.} This yielded a catalog of \tarsnrots\ stars with adopted period measurements.  We show the structure of this catalog in Table~\ref{tab:tars_catalog}.

\subsection{Quality Flagging}\label{subsec:quality_flags}

We created a series of quality flags that can be used to isolate the highest-quality period detections from the catalog above. The name of the flag is followed by its parameter name in Table~\ref{tab:tars_catalog}:

\begin{enumerate}[label={\it \roman*)},leftmargin=12pt,topsep=0pt,itemsep=0ex,partopsep=1ex,parsep=1ex]
    \item \textit{Binary flag} (\texttt{flag\_possible\_binary}) ---
    Boolean flag set \texttt{True} for a target if {\it any} sector used to calculate the adopted period shows a non-harmonic peak (not within 10\% of $nP_{\rm best}$ or $P_{\rm best}/n$ for $n\le 5$) that has power greater than a fixed threshold (default power = 0.2). 
    It will also be set to \texttt{True} if the \gaia\ DR3 RUWE $>1.4$ or \texttt{non\_single\_star}$>0$.
    
    \item \textit{Bad coverage flag} (\texttt{flag\_bad\_coverage}) --- Boolean flag for insufficient temporal coverage relative to the inferred  period. Observation times are split into groups separated by gaps larger than 0.1\,d; the durations of these groups are summed to give an effective coverage window. The flag is set \texttt{True} if this window is less than the period corresponding to the highest peak in the periodogram.
    
    \item \textit{Multiple-period flag}  (\texttt{flag\_multiple\_periods}) --- Boolean flag set to \texttt{True} if any sector period \added{passed the systematics classifier} and lies more than 10\% away from the adopted period and its first-order harmonics ($P_{\rm rot}$, $2P_{\rm rot}$, or $P_{\rm rot}/2$), indicating potential residual multi-period behavior.

    \item \textit{Doubled period flag} (\texttt{flag\_doubled\_period}) --- Boolean flag set to \texttt{True} if any of the sector-level periods involved in calculating the adopted period were doubled.

    \added{
    \item Single passing sector flag (\textit{flag\_one\_cvz\_sector}) --- Boolean flag set to \texttt{True} if period measurements were available for $>=10$ TESS sectors for a given star but only one sector passed the systematics and \added{harmonic} classifiers.
    }
    
    \item \textit{Contamination flag} (\texttt{final\_n\_contams}) --- A nearby star could reproduce the observed variability amplitude ($A_{\rm obs}$) if its variability amplitude ($A_{\rm req}$) is at least
    \begin{equation}
    A_{\rm req} = A_{\rm obs}\,\left(10^{0.4\,\Delta G} + 1\right).
    \end{equation}
    We use $\Delta G$ as a proxy for $\Delta T$ since not all stars needed for this calculation are in the TIC. 
    We assign a maximum plausible intrinsic amplitude ($A_{\rm max}$) for a neighbor, taken to be the 95th percentile of the best fit amplitudes from stars with similar $G_{\rm BP}-G_{\rm RP}$ colors as the neighbor. Any star with $A_{\rm max}>A_{\rm req}$ is taken to be a potential contaminant. The contamination flag is the number of such contaminants within $1\arcmin$ radius (approximately three TESS pixels) of the target in \emph{Gaia} DR3. 
    Since the variability amplitude varies between sectors, we first calculate the number of potential contaminants in each sector, then report the minimum number of contaminants across sectors used to calculate the final period. 
    
\end{enumerate}

If we were to require no possible binaries, no multiple-periods, and no potential contaminating neighbor stars (\texttt{final\_n\_contams = 0}), we would have a catalog of \added{314,650} periods.  Additionally requiring the period to be measured in multiple TESS sectors would yield a catalog of \added{216,007} rotation periods. Based on literature comparisons, these periods are considered the highest quality, but users may readily construct alternative samples or apply their own set of flags based on their science case (see Section~\ref{sec:conclusion} for instructions).

\begin{table*}
    \centering
    \caption{Period measurements for \tarsnrots\ stars in the default TARS catalog.}
    \begin{tabular}{ccc}
    \hline    \hline
    Parameter & Example Value & \textbf{Description} \\
    \hline
    \texttt{TICID} & 273501562 & TESS Input Catalog identifier \\
    \texttt{dr2\_source\_id} & 2078543414254981376 & Gaia DR2 source identifier \\
    \texttt{dr3\_source\_id} & 2078543414254981376 & Gaia DR3 source identifier \\
    \texttt{ra} & 297.726212 & Gaia DR3 right ascension (deg) \\
    \texttt{dec} & 43.067552 & Gaia DR3 declination (deg) \\
    \texttt{parallax} & 3.512931 & Gaia DR3 parallax (mas) \\
    \texttt{pmra} & 0.197835 & Gaia DR3 proper motion in right ascension (mas/yr)\\
    \texttt{pmdec} & -19.28951 & Gaia DR3 proper motion in declination (mas/yr)\\
    \texttt{teff} & 5795.21 & Calculated effective temperature (K) \\
    \texttt{Tmag} & 11.5835 & TESS magnitude (mag) \\
    \texttt{phot\_g\_mean\_mag} & 12.027 & Gaia $G$-band apparent magnitude (mag) \\
    \texttt{phot\_g\_mean\_mag\_0} & 4.641 & Absolute G-band magnitude, corrected for extinction \\
    \texttt{phot\_bp\_mean\_mag} & 12.369 & Gaia $G_{\rm BP}$-band apparent magnitude (mag) \\
    \texttt{phot\_bp\_mean\_mag\_0} & 4.957 & Absolute $G_{\rm BP}$ magnitude, corrected for extinction \\
    \texttt{phot\_rp\_mean\_mag} & 11.517 & Gaia $G_{\rm RP}$-band apparent magnitude (mag) \\
    \texttt{phot\_rp\_mean\_mag\_0} & 4.161 & Absolute $G_{\rm RP}$ magnitude, corrected for extinction \\
    \texttt{BpmRp0} & 0.796 & Extinction-corrected $G_{\rm BP} - G_{\rm RP}$ color (mag) \\
    \texttt{extinction\_a0} & 0.131 & Monochromatic extinction at 550\,nm (mag) \\
    \texttt{ruwe} & 0.932 & Gaia Renormalized Unit Weight Error \\
    \texttt{non\_single\_star} & 0 & Gaia non-single star flag (0 = single, $>$0 = non-single) \\
    \texttt{adopted\_period} & 10.311 & Final adopted rotation period (days) \\
    \texttt{adopted\_period\_unc} & 0.204 & Uncertainty on adopted rotation period (days) \\
    \texttt{flag\_multiple\_periods} & False & Inconsistent periods detected across sectors \\
    \texttt{flag\_possible\_binary} & False & Possible binary based on RUWE, Gaia flags, or light curve \\
    \texttt{flag\_one\_cvz\_sector} & False & 10+ sectors available but only one sector passed the classifiers \\
    \texttt{final\_n\_contams} & 0 & Minimum number of contaminating sources across sectors \\
    \texttt{flag\_doubled\_period} & True & Period doubling was involved in calculating final period \\
    \texttt{n\_secs} & 9 & Number of TESS sectors involved in final period calculation \\
    \texttt{n\_sec\_ratio} & 1.0 & Available-to-used sector ratio for \texttt{adopted\_period} \\
    \texttt{median\_amplitude} & 0.239 & Median of sector amplitudes used to calculate \texttt{adopted\_period} (\%) \\
    \texttt{sectors} & 14,15,41,54,55,74,75,81,82 & Sectors used to calculate \texttt{adopted\_period} \\
    \texttt{sector\_periods} & 10.42,5.42,10.69,9.99, & Sector periods used to calculate \texttt{adopted\_period} (before doubling) \\
    \texttt{} & 10.56,9.62,4.98,5.56,5.14 & \\
    \texttt{sector\_sys\_scores} & 0.01,0.01,0.02,0.00, & Each sector's systematic score \\
    \texttt{} & 0.00,0.00,0.05,0.01,0.10 & \\
    \texttt{sector\_sig\_scores} & 0.99,0.99,0.98,1.00, & Each sector's signal score \\
    \texttt{} & 1.00,1.00,0.95,0.99,0.90 & \\
    \texttt{sector\_match\_scores} & 0.82,0.06,0.92,0.95, & Each sector's match score \\
    \texttt{} & 0.98,0.77,0.00,0.00,0.00 & \\
    \texttt{sector\_harmonic\_scores} & 0.18,0.94,0.09,0.05, & Each sector's harmonic score \\
    \texttt{} & 0.02,0.23,1.00,1.00,1.00 & \\
    \texttt{sector\_match\_probs} & 0.93,0.13,0.95,0.99, & Probability that each sector's period is a match \\
    \texttt{} & 1.00,0.91,0.05,0.04,0.03 & \\
    \texttt{sector\_harmonic\_probs} & 0.07,0.87,0.05,0.01, & Probability that each sector's period is a half-period harmonic \\
    \texttt{} & 0.00,0.09,0.95,0.96,0.97 & \\
    \texttt{sector\_amplitudes} & 0.18,0.15,0.29,0.27, & Amplitude (in percent) of each sector's measurement \\
    \texttt{} & 0.32,0.38,0.12,0.24,0.15 & \\
    \hline
    \multicolumn{3}{c}{\footnotesize \textbf{Note.} This table is published in its entirety in machine-readable format. One entry is shown for guidance regarding form and content.}
    \end{tabular}
    \label{tab:tars_catalog}
\end{table*}

\section{Validation with Literature Periods}\label{sec:validation}

To validate our rotation period measurements, we compare our derived periods against four independent samples with well-established rotation periods: \kepler, \ktwo, ZTF, and an open cluster compilation. Our primary goal is to quantify how the reliability and completeness of our catalog depend on user-defined probability thresholds and quality-flag selections, and to provide guidance on how these choices affect the final sample.

For each dataset, we first consider the simplistic baseline yielded by adopting the highest peak in the Lomb--Scargle periodogram as the rotation period for peaks with Lomb--Scargle power exceeding 0.05.  For this baseline, we also apply the period-adoption logic from Section~\ref{subsec:period_logic}, but without the systematics or \added{harmonic} classifier.  We then compare this baseline against the periods yielded by various classifier thresholds to explore the costs and benefits of our random forest classification. 

\subsection{Kepler}\label{subsec:kepler_val}

For the \kepler\ validation, we calculate our statistics on the holdout validation sample from Section~\ref{subsubsec:alias_rf} and use stars with $T < 16$ and $P_{\rm rot} < 40$ days. This yields a comparison sample of 60,699 sector-level measurements for 27,711 stars. We process
the TESS data for these stars through the same pipeline applied to our TESS targets, including both the systematics (Section~\ref{subsubsec:systematics}) and \added{harmonic} classifiers
(Section~\ref{subsubsec:alias_rf}) with varying random forest probability thresholds. For a given threshold (${\rm RF}=X$), we require that each sector-level measurement passes the systematics and \added{harmonic} classifier (true period or true \added{harmonic}) with the same probability ($>X$). After adopting a final rotation period following Section~\ref{subsec:period_logic}, we compare it to
the \kepler\ period. We repeat this procedure for $X=$ 0.5, 0.6, 0.7, 0.8, and 0.9.

We also consider two quality-flagging regimes: one in which no quality flags (see Section~\ref{subsec:quality_flags}) are applied and another in which \added{we require no contaminants, no binaries, no multiple periods, and that the period was measured in multiple TESS sectors} (the highest quality periods). The results are summarized in Table~\ref{tab:validation_stats}, which reports both the number of recovered stars and the fraction of matches, half-period \added{harmonics}, double-period \added{harmonics}, and non-recoveries. \added{We define double-period harmonics as periods that match twice the true rotation period within $3\sigma$ and non-recoveries as periods that are more than $3\sigma$ away from any of half the true period, the true period, or double the true period.}

In the baseline case with no quality flags applied, only 44.7\% of the adopted periods match the \kepler\ periods, with a substantial fraction (20.8\%) identified as half-period \added{harmonics}. Applying all quality flags \added{(i.e. requiring no possible binaries, no multiple periods, no contaminants, and that the period was measured in more than one sector)} improves reliability: the match fraction increases to 68.8\%, albeit at the cost of completeness, with periods recovered for $\sim3\%$ of the validation sample.

A similar tradeoff is observed when varying the classifier probability thresholds. Without quality flags, increasing the threshold from 0.5 to 0.9 raises the match fraction from \added{66.1}\% to \added{77.8}\%. When all quality flags are applied, the corresponding match fractions increase from \added{85.5}\% to \added{89.4}\%. In all cases, higher thresholds suppress \added{harmonics} but reduce the total number of retained stars.

In terms of \added{harmonic} contamination, the half-period \added{harmonic} rate decreases from 20.8\% in the un-flagged baseline sample to just \added{0.7}\% when adopting strict probability thresholds ($>0.9$) and applying all quality flags. For users seeking an especially clean sample, requiring a threshold $>0.9$ from both the systematics and \added{harmonic} classifiers (i.e.\ the probability that a period is the true period $>0.9$) provides a clean sample. Under these criteria, and with all quality flags applied, we recover \added{680} Kepler stars with match, half-\added{harmonic}, double-\added{harmonic}, and non-recovery fractions of \added{0.894}, \added{0.007}, \added{0.000}, and \added{0.099}, respectively. Without quality flags, the same selection yields \added{4,694} stars with corresponding fractions of \added{0.778}, \added{0.026}, \added{0.013}, and \added{0.184}.

\begin{figure*}
    \centering
    \includegraphics[width=0.9\linewidth]{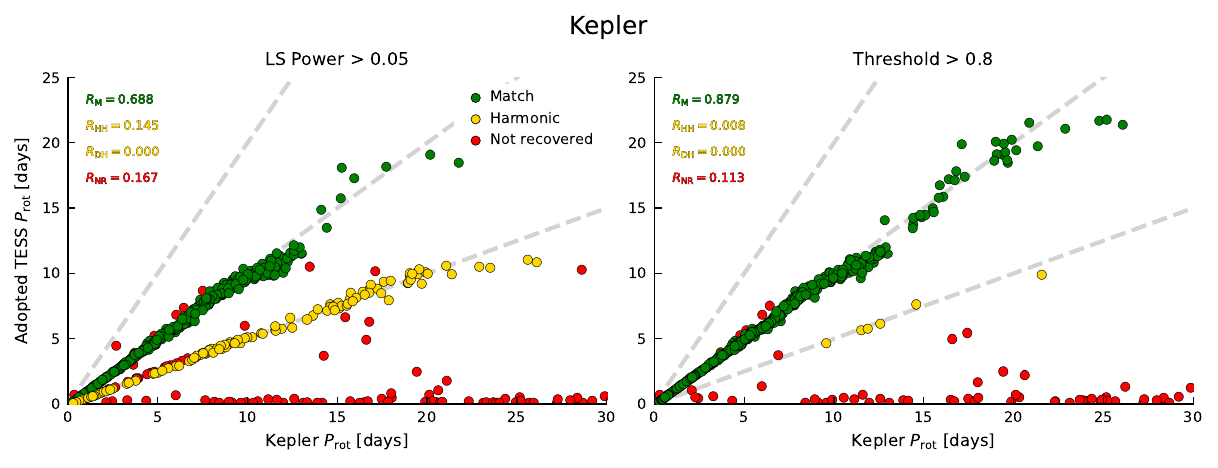}
    \includegraphics[width=0.9\linewidth]{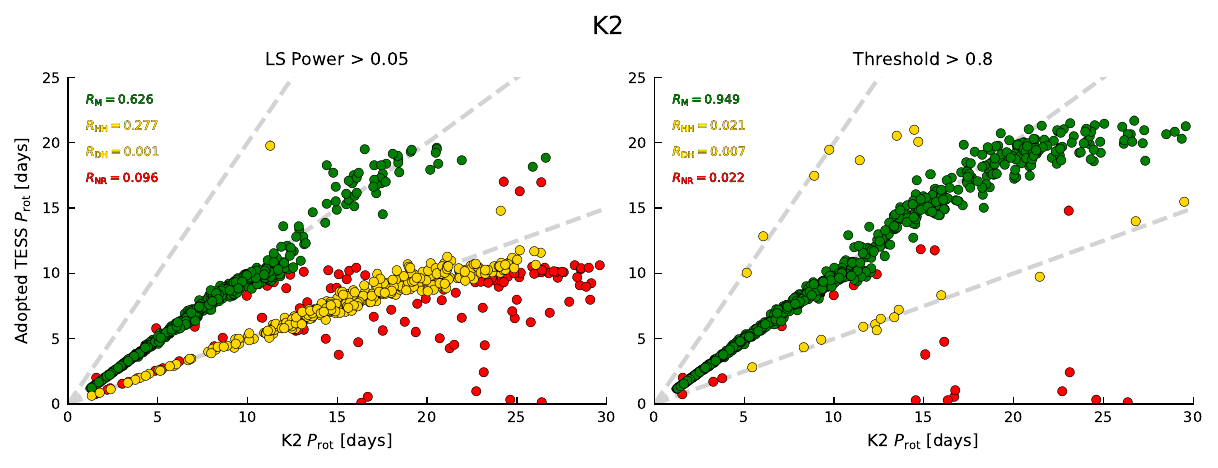}
    \includegraphics[width=0.9\linewidth]{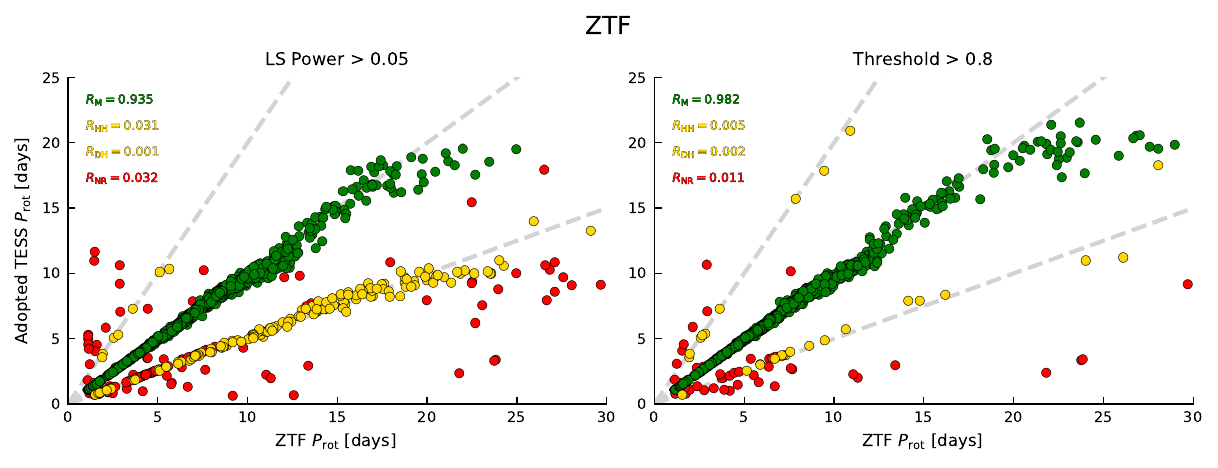}
    \caption{
    \textbf{Validation of our vetting procedure using \kepler, \ktwo, and ZTF.}
    \textit{Top}: \kepler\ comparison to \citet{mcquillanROTATIONPERIODS342014}. \textit{Middle}: \ktwo\ comparison to \citet{reinholdStellarRotationPeriods2020}. \textit{Bottom}: ZTF comparison to \citet{Lu2022}. Each panel lists the rates of matches ($R_{\rm M}$), half-period \added{harmonics} ($R_{\rm HH}$), double-period \added{harmonics} ($R_{\rm DH}$), and non-recoveries ($R_{\rm NR}$).
    The left panels show a baseline approach adopting the highest Lomb--Scargle peak (LS power $>0.05$). Green points denote $3\sigma$ matches, yellow indicate \added{harmonics}, and red are non-recoveries; a prominent half-period \added{harmonic} sequence is evident in all surveys.
    The right panels show results after applying our full vetting procedure, including systematics and \added{harmonic} random-forest classifiers and all quality flags. The half-period sequence is largely reduced, yielding agreement across a broad period range. Residual discrepancies are dominated by unidentified \added{harmonics} (0.8\% in \kepler, 2\% in \ktwo, 0.5\% in ZTF) and doubled periods (0\% in \kepler, 0.7\% in \ktwo, 0.2\% in ZTF). Overall, the method reliably recovers rotation periods out to $P_{\rm rot}\sim25$~days, beyond the typical single-sector TESS limit.
    }
    \label{fig:kepler_k2_before_after}
\end{figure*}

\added{These results are shown in the top row of Figure~\ref{fig:kepler_k2_before_after}. The left column shows the baseline validation set and the right column shows the validation set after applying the TARS pipeline (in both plots we we require no contaminants, no binaries, no multiple periods, and that the period was measured in multiple TESS sectors). Our systematics and \added{harmonic} classifiers largely remove spurious detections and correct the half-period \added{harmonics} seen in the left column, but $\sim0.8\%$ of half-period \added{harmonics} are missed by our \added{harmonic} classifier.
}

\subsection{K2}\label{subsec:k2_val}

We perform an analogous validation using rotation periods measured from the \ktwo\ mission. For our \ktwo\ comparison sample, we take the rotation period measurements from
\citet{reinholdStellarRotationPeriods2020} with $P_{\rm rot} < 40$~days and require $H_{\rm peak} > 0.3$. After cross-matching with our full target list, we are left with 16,823 stars in common. As for the
\kepler\ validation, we process these stars through the full TARS pipeline and adopt final periods using the same logic described in Section~\ref{subsec:period_logic}.

For the baseline \added{Lomb-Scargle power $>0.05$} comparison, only 35.6\% of the adopted periods match the published \ktwo\ periods, with a large fraction (32.9\%) identified as half-period \added{harmonics} (Table~\ref{tab:validation_stats}). This
highlights the prevalence of \added{half-period harmonics} in \ktwo\ light curves when no additional vetting is applied.

\added{Applying the random forest classifier vetting methodology developed here} substantially improves the reliability of the recovered periods. Without quality flagging, increasing the probability threshold from 0.5 to 0.9 raises the match fraction from \added{75.3}\% to \added{86.5}\%, while reducing the half-period \added{harmonic} rate from \added{9.7}\% to \added{5.1}\%. When all quality flags are enforced, the match fraction improves further, reaching \added{96.1}\% at a threshold of
0.9, with a corresponding half-period \added{harmonic} rate of \added{1.7}\%. As expected, these gains in reliability come at the cost of completeness: the number of recovered stars decreases from 1,319 in the baseline \added{sample with quality flags applied (no contaminants, no binaries, no multiple periods, period was measured in multiple TESS sectors)} to \added{870} at the most stringent threshold.

These results are shown visually in the middle row of Figure~\ref{fig:kepler_k2_before_after}. The trends mirror those seen for \kepler: the vast majority of stars that pass our vetting agree with the
\ktwo\ period, with \added{$\sim0.7\%$} of the remaining measurements being inadvertently doubled. As in the \kepler\ case, our \added{harmonic} flagging and period-doubling logic enables recovery of rotation periods beyond
what is typically achievable in a standard single-sector TESS light curve.

\begin{sidewaystable*}[p]
\centering
\caption{Validation statistics for TARS rotation period recovery across different samples and probability thresholds.}
\label{tab:validation_stats}
\begin{tabular}{llccccccccccc}
\toprule
& & \multicolumn{5}{c}{No Quality Flags} & \multicolumn{5}{c}{All Quality Flags} \\
\cmidrule(lr){3-7} \cmidrule(lr){8-12}
Sample & Threshold & $N_\mathrm{stars}$ & Match & Half Harmonic & Double Harmonic & Non-recov. & $N_\mathrm{stars}$ & Match & Half Harmonic & Double Harmonic & Non-recov. \\
\midrule
Kepler ($N_{\rm tot}$ = 27,711) & Baseline & 8097 & 0.447 & 0.208 & 0.007 & 0.338 & 863 & 0.688 & 0.145 & 0.000 & 0.167 \\
     & 0.5 & 8660 & 0.661 & 0.059 & 0.048 & 0.231 & 934 & 0.855 & 0.017 & 0.010 & 0.118 \\
     & 0.6 & 7602 & 0.697 & 0.046 & 0.036 & 0.221 & 883 & 0.857 & 0.015 & 0.007 & 0.121 \\
     & 0.7 & 6760 & 0.724 & 0.038 & 0.029 & 0.209 & 825 & 0.869 & 0.013 & 0.005 & 0.113 \\
     & 0.8 & 5905 & 0.745 & 0.031 & 0.022 & 0.203 & 780 & 0.879 & 0.008 & 0.000 & 0.113 \\
     & 0.9 & 4694 & 0.778 & 0.026 & 0.013 & 0.184 & 680 & 0.894 & 0.007 & 0.000 & 0.099 \\
\midrule
K2 ($N_{\rm tot}$ = 16,823) & Baseline & 5980 & 0.356 & 0.329 & 0.005 & 0.311 & 1319 & 0.626 & 0.277 & 0.001 & 0.096 \\
     & 0.5 & 5533 & 0.753 & 0.097 & 0.035 & 0.115 & 1482 & 0.914 & 0.038 & 0.023 & 0.025 \\
     & 0.6 & 4990 & 0.784 & 0.083 & 0.032 & 0.101 & 1328 & 0.927 & 0.030 & 0.020 & 0.023 \\
     & 0.7 & 4506 & 0.811 & 0.072 & 0.028 & 0.089 & 1216 & 0.938 & 0.021 & 0.015 & 0.026 \\
     & 0.8 & 3956 & 0.836 & 0.064 & 0.023 & 0.077 & 1069 & 0.949 & 0.021 & 0.007 & 0.022 \\
     & 0.9 & 3127 & 0.865 & 0.051 & 0.019 & 0.066 & 870 & 0.961 & 0.017 & 0.003 & 0.018 \\
\midrule
Clusters ($N_{\rm tot}$ = 2,632) & Baseline & 2022 & 0.727 & 0.091 & 0.012 & 0.169 & 599 & 0.853 & 0.077 & 0.010 & 0.060 \\
       & 0.5 & 2147 & 0.826 & 0.017 & 0.043 & 0.114 & 645 & 0.924 & 0.012 & 0.029 & 0.034 \\
       & 0.6 & 2089 & 0.831 & 0.017 & 0.045 & 0.108 & 634 & 0.926 & 0.011 & 0.028 & 0.035 \\
       & 0.7 & 2005 & 0.842 & 0.017 & 0.037 & 0.103 & 603 & 0.937 & 0.012 & 0.015 & 0.036 \\
       & 0.8 & 1905 & 0.851 & 0.014 & 0.033 & 0.102 & 573 & 0.949 & 0.007 & 0.009 & 0.035 \\
       & 0.9 & 1746 & 0.871 & 0.012 & 0.027 & 0.090 & 548 & 0.956 & 0.000 & 0.009 & 0.035 \\
\midrule
ZTF ($N_{\rm tot}$ = 18,894) & Baseline & 11905 & 0.807 & 0.076 & 0.004 & 0.113 & 4774 & 0.935 & 0.031 & 0.001 & 0.032 \\
       & 0.5 & 11920 & 0.895 & 0.024 & 0.018 & 0.063 & 5092 & 0.974 & 0.006 & 0.006 & 0.014 \\
       & 0.6 & 11467 & 0.910 & 0.021 & 0.013 & 0.056 & 4994 & 0.977 & 0.005 & 0.004 & 0.014 \\
       & 0.7 & 11013 & 0.923 & 0.018 & 0.009 & 0.050 & 4864 & 0.979 & 0.005 & 0.003 & 0.013 \\
       & 0.8 & 10505 & 0.935 & 0.015 & 0.007 & 0.043 & 4723 & 0.982 & 0.005 & 0.002 & 0.011 \\
       & 0.9 & 9663 & 0.951 & 0.008 & 0.004 & 0.037 & 4435 & 0.988 & 0.002 & 0.001 & 0.009 \\
\bottomrule
\end{tabular}
\tablecomments{
Column descriptions are as follows.
(1) \textbf{Sample}: External validation dataset used for comparison, including \kepler, \ktwo, ZTF, and open clusters.
(2) \textbf{Threshold}: Probability threshold applied to both the systematics and \added{harmonic} random forest classifiers; only sector-level measurements with classifier probabilities exceeding this value are
retained.
(3) \textbf{Quality Flags}: Indicates whether the full set of quality flags described in Section~\ref{subsec:quality_flags} (multi-sector agreement, good temporal coverage, no strong evidence for binarity
or contamination, more than one sector used in period calculation) is applied.
(4) \textbf{$N_{\rm stars}$}: Number of stars for which a final adopted rotation period is recovered under the specified selection criteria.
(5) \textbf{Match}: Fraction of recovered stars whose adopted TESS rotation period agrees with the literature period within $3\sigma$, as defined by Equation~1 of \citet{Boyle2025a}.
(6) \textbf{Half-harmonic}: Fraction of recovered stars whose adopted period matches half of the literature rotation period within $3\sigma$, corresponding to $P_{\rm rot}/2$ \added{harmonics}.
(7) \textbf{Double-harmonic}: Fraction of recovered stars whose adopted period matches twice the literature rotation period within $3\sigma$, typically arising from incorrect harmonic correction.
(8) \textbf{Not recovered}: Fraction of stars in the validation sample for which no rotation period satisfying the specified criteria is recovered.
}
\end{sidewaystable*}

\subsection{ZTF}\label{subsec:ztf_val}

For the Zwicky Transient Facility (ZTF) comparison, we adopt rotation periods from \citet{Lu2022}. After cross-matching with our target list, we obtain a validation sample of 18,894 stars that satisfy our selection criteria.

In the baseline case with no quality flags applied, 80.7\% of the adopted periods match the ZTF literature values, with 7.6\% identified as half-period \added{harmonics} and 11.3\% classified as non-recoveries
(Table~\ref{tab:validation_stats}). Applying all quality flags increases the match fraction to 93.5\%, demonstrating that our periods are generally consistent with ZTF-derived period
measurements.

Increasing the random forest probability threshold further improves reliability. Without quality flags, the match fraction rises from \added{89.5}\% at a threshold of 0.5 to \added{95.1}\% at a threshold of 0.9. When all
quality flags are enforced, the match fraction increases from \added{97.4}\% to \added{98.8}\% over the same range, while the half-period \added{harmonic} rate decreases to \added{0.2}\%. As in the other validation samples, these gains come
at the cost of completeness, with the number of recovered stars decreasing from 4,774 in the baseline flagged sample to \added{4,435} at the most stringent threshold.

The ZTF validation results are shown in the bottom row of Figure~\ref{fig:kepler_k2_before_after}. Compared to \kepler\ and \ktwo, ZTF exhibits a lower baseline \added{harmonic} rate and higher overall agreement,
likely due to the ZTF sample being dominated by M dwarfs that inherently have higher rotation amplitudes. Nevertheless, the same reliability--completeness tradeoffs apply, and the TARS framework provides a
consistent and flexible means of selecting rotation samples with well-quantified purity across space-based and ground-based surveys.

\subsection{Open Clusters}\label{subsec:cluster_val}

Cluster membership provides an independent and physically motivated benchmark, as stars of similar age are expected to occupy coherent loci in rotation–effective temperature space. We adopt rotation
periods for eight open clusters from \citet{2023ApJS..268...30L} that are well-represented in our survey: Taurus (4 Myr), $\rho$~Ophiuchus (9 Myr), Upper Scorpius (19 Myr), the Pleiades (123 Myr), NGC~1750
(316 Myr), Praesepe (741 Myr), the Hyades (758 Myr), and NGC~6774 (2.9 Gyr). Together, these clusters span nearly three orders of magnitude in age, providing a complementary validation sample to the
\kepler, \ktwo, and ZTF comparisons presented above.

This cluster validation sample contains 2,632 stars that satisfy our selection criteria. As above, we process these stars through the full TARS pipeline and evaluate recovery statistics as a function of
random forest probability threshold and quality-flag selection.

In the baseline case with no quality flags applied, 72.7\% of the adopted periods match the literature values, with 9.1\% identified as half-period \added{harmonics} and 16.9\% classified as non-recoveries
(Table~\ref{tab:validation_stats}). Compared to the field-star samples, the baseline cluster match fraction is substantially higher, reflecting the coherence of cluster rotation sequences and the
higher-quality light curves available for many of these targets.

Applying increasingly stringent probability thresholds improves the match fraction in a manner similar to the \kepler\ and \ktwo\ samples. Without quality flags, the match fraction increases from \added{82.6}\% at
a threshold of 0.5 to \added{87.1}\% at a threshold of 0.9. When all quality flags are applied \added{(no contaminants, no binaries, no multiple periods, period was measured in multiple TESS sectors)}, the match fraction rises from \added{92.4}\% to \added{95.6}\% over the same threshold range. Selecting stars with \added{$S_{\rm signal} > 0.9$} and $P(\rm match) > 0.9$ or $P(\rm harmonic) > 0.9$ almost entirely eliminates half-period \added{harmonics}, reducing their incidence to $0.0\%$. As before, these improvements in reliability are accompanied by a reduction in completeness,
with the number of retained stars decreasing from 599 in the baseline flagged sample to \added{548} at the most stringent threshold.

\begin{figure*}
    \centering
    \includegraphics[width=1\linewidth]{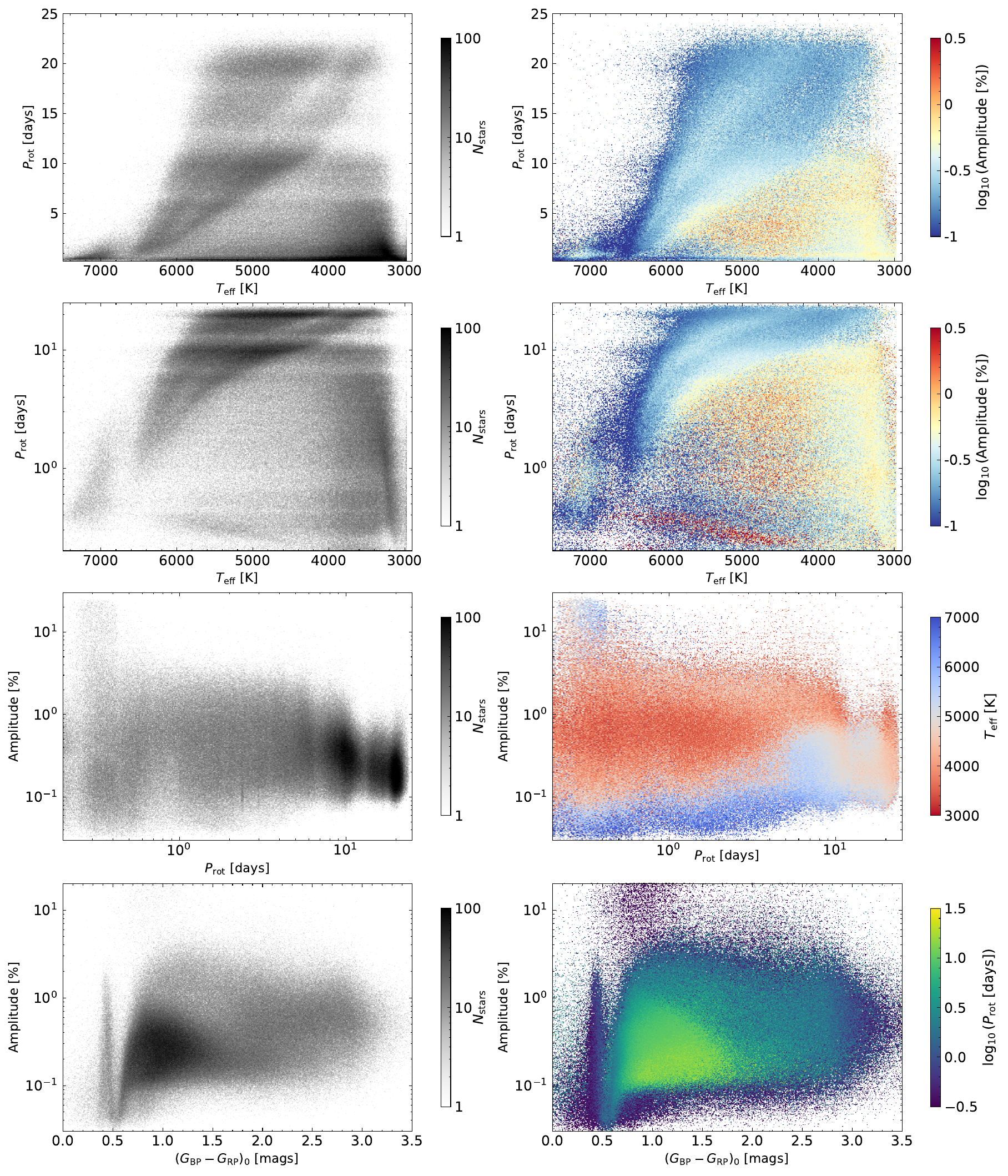}
    \caption{
    \textbf{The default TARS sample as functions of period, temperature, and amplitude.}
    \textit{Top two rows:} Rotation-effective temperature diagrams with linear-scaled $P_{\rm rot}$ (top row) and log-scaled $P_{\rm rot}$. The left panel presents the number density of stars across the sample, while the right panel colors each bin by the median amplitude of the rotators in that bin.
    \textit{Middle row:} Rotation-amplitude diagram with log-scaled $P_{\rm rot}$. The bins in the right column are colored by median effective temperature.
    \textit{Bottom row:} Color-amplitude diagram colored by number of stars on the left and $P_{\rm rot}$ on the right. The structure in these plots is discussed in Section~\ref{sec:discussion}.
    }
    \label{fig:rot_amp_cmd}
\end{figure*}

\begin{figure*}
    \centering
    \includegraphics[width=1\linewidth]{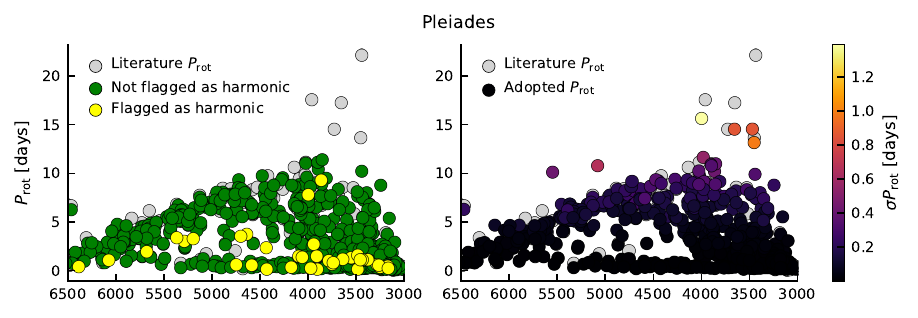}
    \includegraphics[width=1\linewidth]{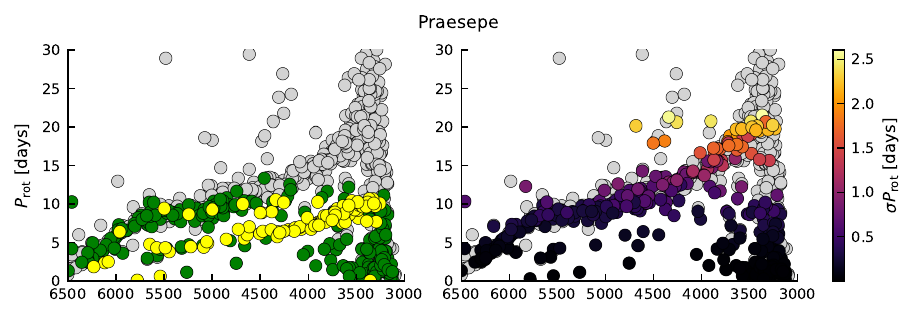}
    \includegraphics[width=1\linewidth]{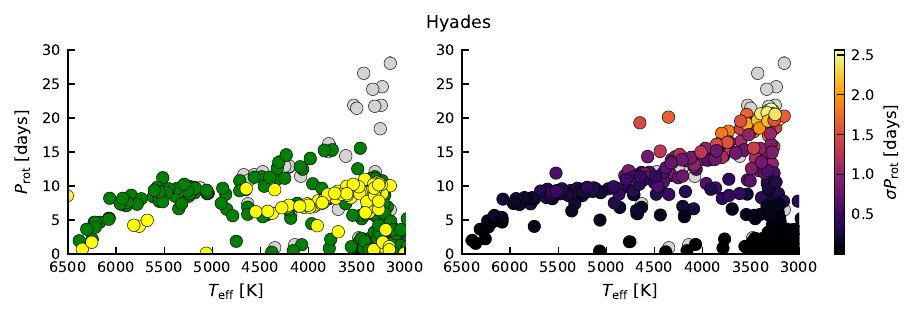}
    \caption{
    \textbf{Our harmonic classifier enables reconstruction of open cluster rotation-temperature sequences.}
    \textit{Left column:} Rotation period–effective temperature sequence for the Pleiades ($t \sim 120$ Myr), Praesepe ($t \sim 670$ Myr), and the Hyades ($t \sim 700$ Myr). We adopt membership lists for each cluster from \citet{Hunt2023}. Literature rotation periods are taken from \citet{rebullROTATIONPLEIADESDATA2016} for the Pleiades, \citet{rampalliThreeK2Campaigns2021} for Praesepe, and \citet{douglasK2RotationPeriods2019} for the Hyades and are marked in gray. Green points indicate TESS measurements with $P(\rm match) > 0.8$ and yellow points denote measurements with $P(\rm harmonic) > 0.8$. A clear secondary sequence at approximately half the literature period is visible in Praesepe and the Hyades for $3500 \lesssim T_{\rm eff} \lesssim 6000$~K, and the majority of stars on this sequence are correctly identified as \added{harmonics}.
    \textit{Right column:} The adopted periods for each of these stars from the default TARS catalog, colored by the uncertainty on the period. Our harmonic flagging and doubling procedure allows us to correct \added{harmonics} and recover rotation periods that closely reproduce the literature rotation sequences.
    }
    \label{fig:praesepe_alias}
\end{figure*}

\begin{figure*}
    \centering
    \includegraphics[width=1\linewidth]{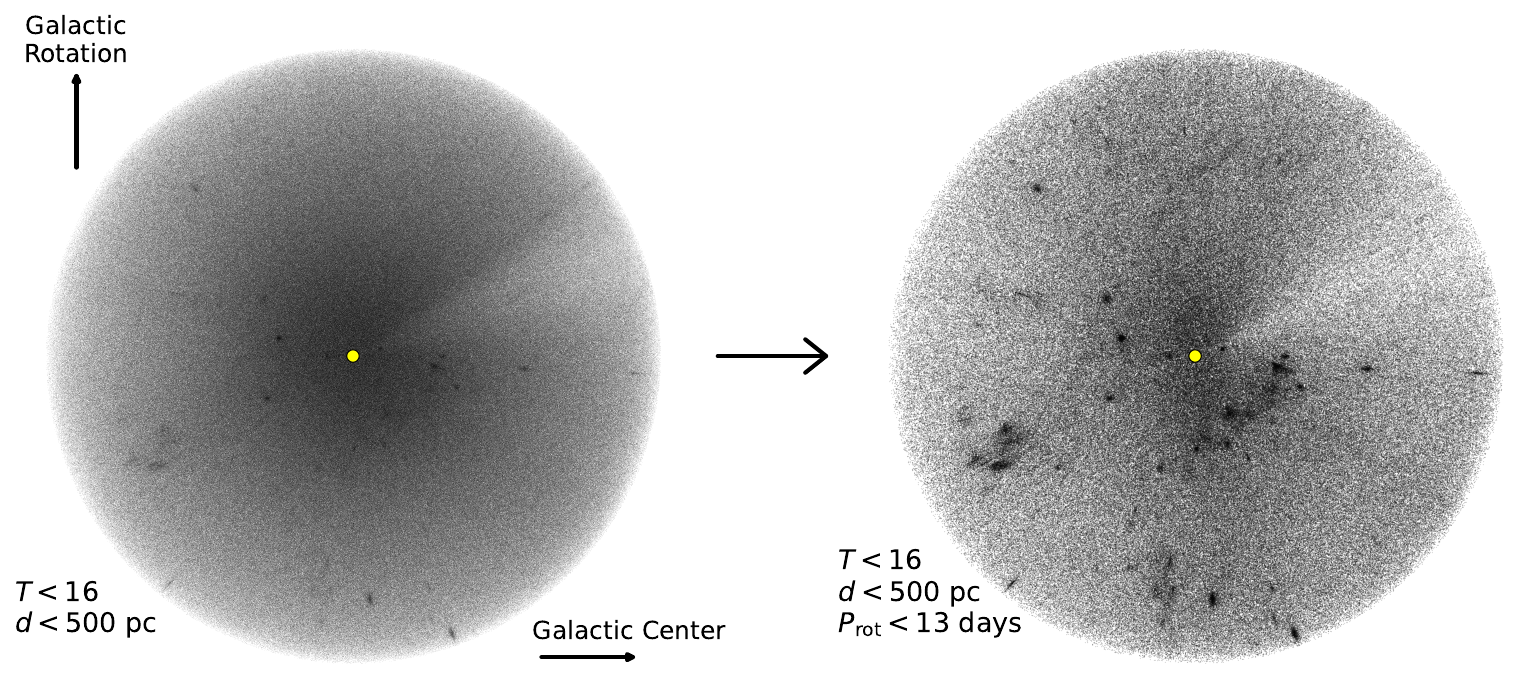}
    \caption{
    \textbf{Selecting fast rotators ($P_{\rm rot} < 13$~days) enhances the visibility of clustered populations.}
    \textit{Left:} Distribution in Galactocentric X and Y coordinates of all stars in the TARS target list satisfying $T < 16$ and $d < 500$~pc. The Sun is marked as a yellow point at the origin. 
    \textit{Right:} The same sky region, restricted to stars with adopted rotation periods $P_{\rm rot} < 13$~days. After applying this cut, clustered populations become more prominent.  An interactive version is {\bf \href{https://lgbouma.com/tars_viz/}{available online}}.
    }
    \label{fig:xy_before_after}
\end{figure*}

\section{Results}\label{sec:results}

\subsection{Overall Catalog Properties}
The TARS catalog provides period estimates for \tarsnrots\ stars within 500\,pc. If we apply our suggested quality cuts (Section~\ref{subsec:quality_flags}) to remove potential contaminants, targets with evidence of multi-period behavior, likely binaries, and stars that share a period with a nearby star, we have a `clean' sample of \added{314,650} stars. Of these, \added{216,007} stars constitute our highest-quality subset, possessing consistent period measurements across multiple TESS sectors. We show the distribution of the \tarsnrots\ star sample as a function of period, amplitude and effective temperature in Figure~\ref{fig:rot_amp_cmd}. We examine structure in these diagrams in the discussion (Section~\ref{sec:discussion}) below.

Our default catalog reports rotation periods for \added{50,816} stars with $T < 16$ and $d < 100$~pc. To assess how many of these periods are newly measured, we cross-matched our target list against a set of previously published rotation period catalogs. For each catalog, we list the number of matches with $T < 16$ and $d < 100$~pc, along with the corresponding instrument used to measure the rotation periods: \citet{Kounkel2022} (TESS; 1,782 stars), \citet{2023ApJS..268....4F} (TESS; 12,632), \citet{2026ApJS..282...10S} (ASAS-SN; 1,000), \citet{Lu2022} (ZTF; 2,419), \citet{2022ApJ...936..138H} (TESS; 4,670), \citet{2024AJ....167..189C} (TESS; 3,454), \citet{Claytor_2024} (TESS; 685), \citet{reinholdStellarRotationPeriods2020} (\ktwo; 1,094), \citet{mcquillanROTATIONPERIODS342014} (\kepler; 103), \citet{2018AJ....155...39O} (KELT; 5,233), \citet{Newton2016} (MEarth; 230), \citet{2018AJ....156..217N} (MEarth; 188), \citet{2023ApJS..268...30L} (\kepler\ and \ktwo; 211), \citet{Jayasinghe2018} (ASAS-SN; 850), \citet{Van-Lane2025} (multiple instruments; 159), and \citet{2023A&A...674A..20D} (Gaia DR3; 1,867). Together, these sources provide prior period measurements for 27,437 stars in the $T < 16$, $d < 100$~pc sample. Our catalog contains \added{36,246} stars with $T < 16$ and $d < 100$ pc that do not have reported periods in the above catalogs. Our catalog therefore increases the number of bright stars with known rotation periods within 100~pc by a factor of \added{2.3}. Applying the same calculation out to 500~pc yields an increase by a factor of \added{4.0}. The spatial distribution of the catalog viewed top-down in Galactocentric coordinates is shown in Figure~\ref{fig:xy_before_after}, where selecting fast rotators enhances the visibility of young stellar associations.

\subsection{Overall Reliability of the TARS catalog}

The validation tests presented in Section~\ref{sec:validation} imply that the default TARS catalog is reliable across a wide range of stellar populations. When adopting our default thresholds (systematics \added{score} and \added{harmonic} classifier probabilities $>0.8$) without imposing additional quality flags, the recovered rotation periods agree with independent literature measurements at the $\sim$85–90\% level, depending on the comparison sample. Agreement is highest for ZTF and open clusters, where baseline match fractions exceed 80\%, and somewhat lower for the \kepler\ and \ktwo\ samples, which are more susceptible to half-period \added{harmonics} in the absence of our additional vetting. Averaged over all validation sets and accounting for the relative sample sizes, we estimate that approximately $\sim85\%$ of the periods in the default catalog correspond to the true stellar period, $\sim5\%$ half-period \added{harmonics}, $\sim5\%$ double-period \added{harmonics}, and $\sim5\%$ blends or contamination from nearby stars.

When requiring multi-sector agreement, good temporal coverage, no evidence of binarity or multi-period behavior, and no plausible contaminating neighbors, the resulting subset exhibits match fractions of $\gtrsim85\%$ across all validation samples, reaching $\gtrsim90\%$ for open clusters and ZTF and $\sim88\%$ for \kepler. Under these stricter criteria, we estimate that $\sim92\%$ of these measurements reflect the true rotation period. Residual contamination is dominated by a small fraction of doubled periods rather than half-period \added{harmonics}, and the incidence of non-matches is correspondingly low.

We took the additional step of inspecting some of the non-matches across the various samples. Using the \kepler\ 0.8 threshold with quality flags on from Table~\ref{tab:validation_stats}, we find 63 stars where our measurement did not agree with the \kepler\ measurement. After visually examining these measurements, we find that our measurements typically agree with the periodic signal present in our light curves. For the $\sim9\%$ non-match rate in the \kepler\ validation sample, we find that $76\%$ of this contamination comes from stars with adopted periods of less than one day. 

One explanation for the poorer \kepler\ recovery is blending or contamination: the median distance to the stars with rotation periods that were not recovered in the \kepler\ sample is 886~pc, compared to 579~pc for the stars with rotation periods that were recovered. Comparatively, the median distance to both the matches and non-recoveries in the \ktwo\ sample is $\sim190$~pc. For a fixed angular resolution (the 21\arcsec{} TESS pixel), the probability of a background source contaminating the aperture increases with the distance probed through the Galactic disk. Furthermore, the \kepler{} field’s relatively low Galactic latitude ($b\simeq13\deg$) results in a higher integrated background stellar density. This increased crowding both dilutes the signal from the primary and risks introducing other periodic signals from nearby stars, including those with amplitudes exceeding those captured by our contamination flag (e.g., eclipsing binaries). Indeed, when we manually inspected the difference images (rotation peak-trough) of the TESS pixel files, many were centered on an adjacent pixel. Fortunately, this does not seem to be an issue for the main catalog as the stars are much closer ($d<500$\,pc) and spread over the whole sky.

Overall, these results show that the reliability of the TARS catalog is both high and tunable. Users interested in maximizing sample size can work with the default catalog while accepting a modest level of contamination, whereas users requiring highly secure rotation periods --- such as gyrochronology calibration, cluster membership analysis, or detailed comparisons to stellar evolution models --- can readily isolate subsets of the data with reliability approaching that of space-based long-baseline surveys.

\subsection{Effectiveness of Harmonic-Flagging}

The effectiveness of our \added{harmonic}-flagging procedure is illustrated in Figure~\ref{fig:praesepe_alias}. In this figure, we cross match membership lists for the Pleiades, Praesepe, and the Hyades from \citet{Hunt2023} with the default TARS catalog with quality flags applied. For reference, we show literature K2 rotation sequences for the Pleiades \citep{rebullROTATIONPLEIADESDATA2016}, for Praesepe \citep{rampalliThreeK2Campaigns2021}, and for the Hyades \citep{douglasK2RotationPeriods2019}. 

In the left column, measurements flagged as \added{harmonics} with $P(\rm harmonic)>0.8$ are shown in yellow, literature periods in gray, and measurements flagged as matches with $P(\rm match) > 0.8$ are in green. At higher effective temperatures our recovered periods closely follow the literature sequence. Toward cooler temperatures and longer periods, an increasing fraction of recovered periods fall systematically below the literature sequence, forming a distinct lower sequence that is characteristic of half-period \added{harmonics}. This is especially true for Praesepe and the Hyades. 

Most stars on these lower sequences are flagged by our \added{harmonic} classifier. The right panel shows the recovered rotation sequences after applying our doubling logic. In particular, this procedure recovers the slow-rotator sequence at periods $\gtrsim 13$~days, which is often assumed to be challenging to measure reliably from single-sector TESS observations. The cost is that a small number of stars were incorrectly identified as \added{harmonics}, pushing them well above the sequence. This can be mitigated by adjusting the \added{harmonic} classifier threshold \added{and applying quality flags}, at the cost of losing some correctly-classified periods (Table~\ref{tab:validation_stats}).

\begin{figure*}
    \centering
    \includegraphics[width=1\linewidth]{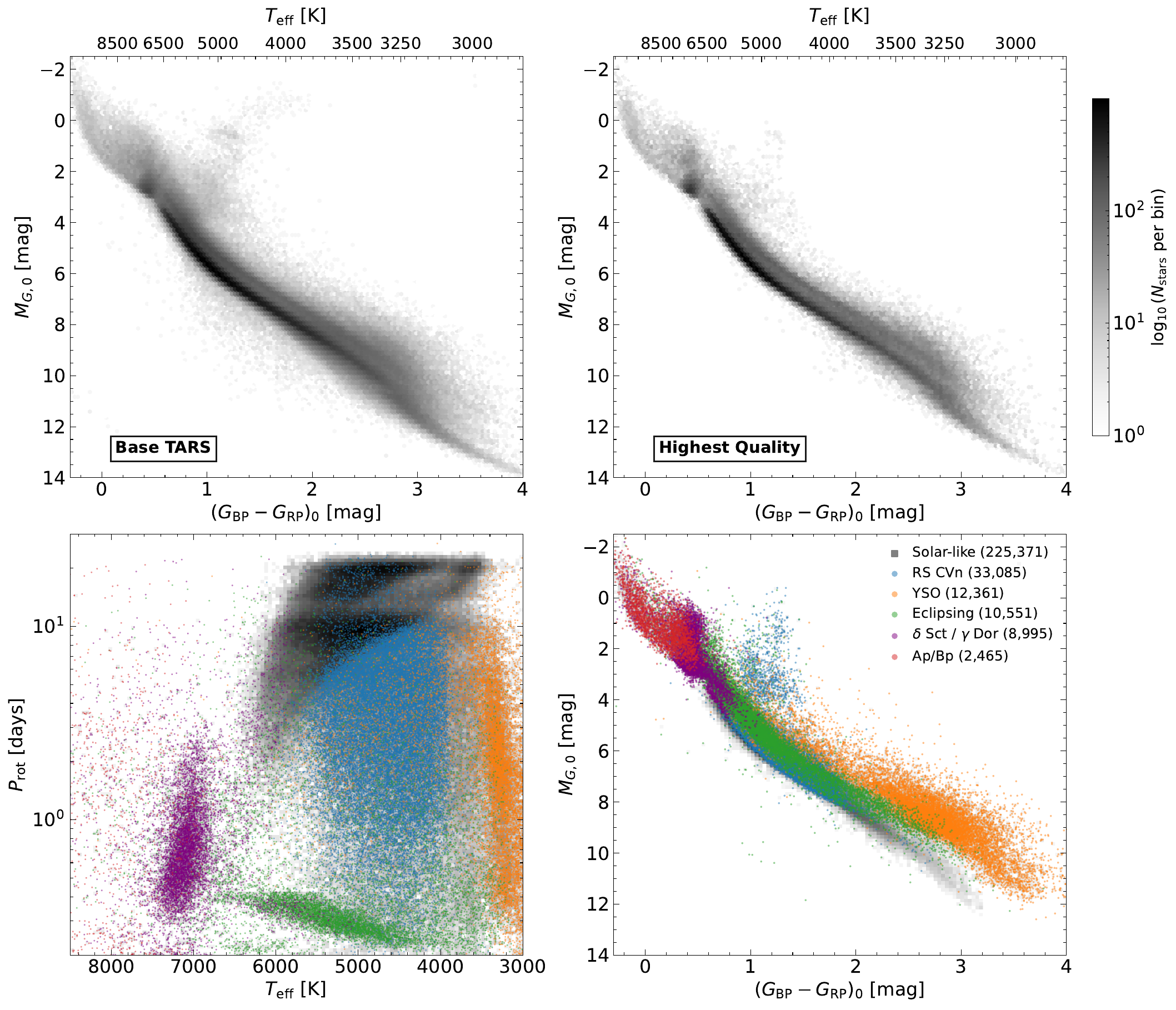}
    \caption{\textbf{Gaia DR3 variability classifications in period--temperature and color--magnitude space.}
    \textit{Left:} Period versus effective temperature for stars in the TARS catalog cross-matched with the \gaia\ DR3 \texttt{vari\_classifier\_result} table. \textit{Right:} Dereddened color--magnitude diagram for the same sources. Points are colored by Gaia variability class: Solar-like modulation (gray), RS CVn systems (blue), young stellar objects (pink), eclipsing binaries (orange), $\delta$ Scuti / $\gamma$ Doradus pulsators (green), and chemically peculiar Ap/Bp stars (yellow). The number of stars in each class is listed in the legend. The strong dominance of rotational classes supports the conclusion that the vast majority of measured TARS periods correspond to stellar rotation.}
    \label{fig:variability}
\end{figure*}

Table~\ref{tab:validation_stats} shows that the completeness for clusters in our default catalog should be \added{$\sim72\%$}. An independent test of this estimate follows from noting that 735 Hyads from \citet{Hunt2023} overlap with the TARS target list, and \added{498} of these stars appear in our default catalog.  These counts suggest a completeness of \added{68\%}, in close alignment with the expected value. Praesepe is further away ($d\sim190$ pc compared to $\sim50$ pc for the Hyades), implying a more challenging test case. \citet{Hunt2023} lists 803 Praesepe members that overlap with the default TARS catalog and \added{391} appear in the default catalog, implying a completeness of \added{49\%}.

\begin{figure*}
    \centering
    \includegraphics[width=1\linewidth]{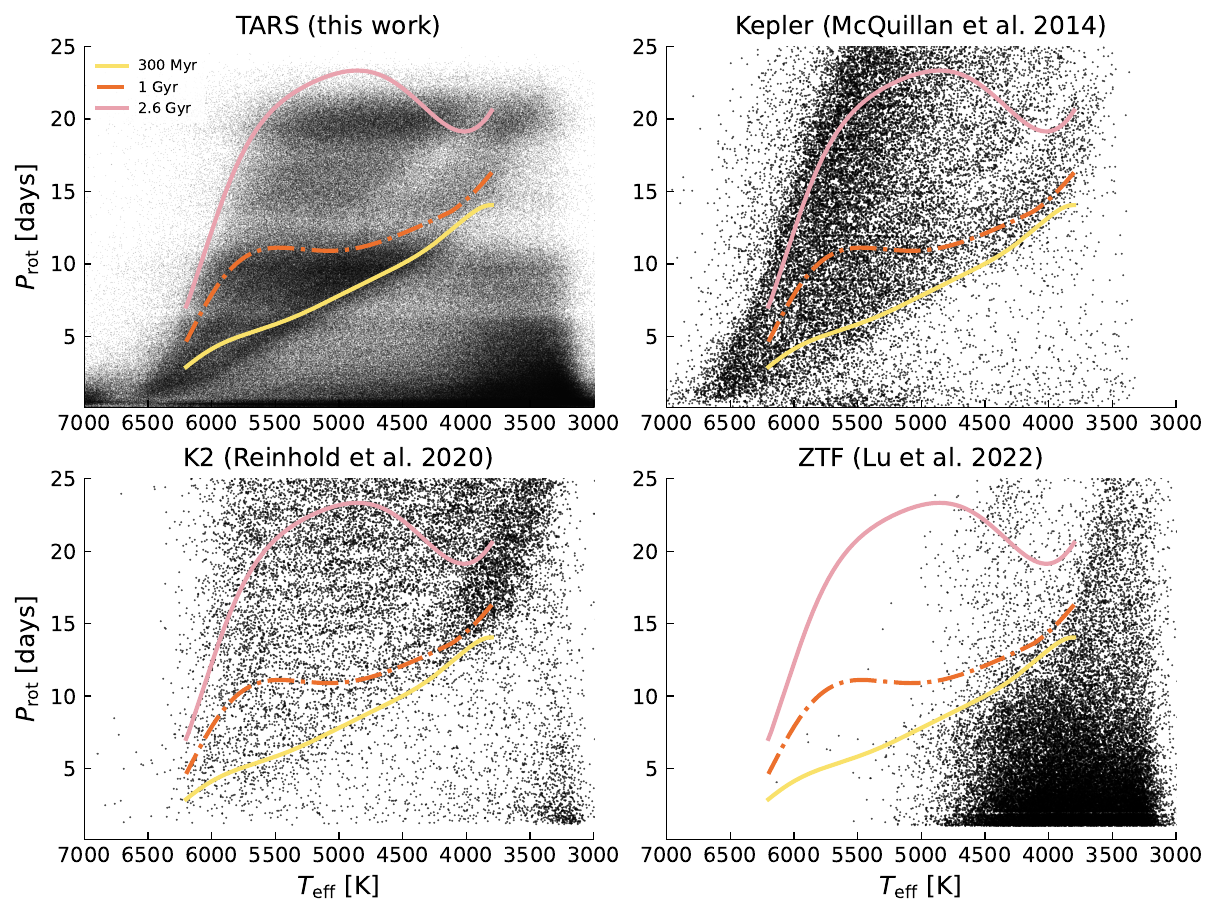}
    \caption{
    \textbf{The TARS \prot-\teff\ plane recovers structures seen in longer baseline surveys.}
    \textit{Top left:} The default TARS catalog with rotational isochrones from NGC-3532 \citep{Fritzewski2021}, NGC-6811 \citep{curtisTemporaryEpochStalled2019}, and Ruprecht-147 \citep{Curtis2020} overplotted. The structure in this diagram is discussed in Section~\ref{subsec:prot_teff_structure}.
    \textit{Top right:} The sample of \kepler\ rotators from \citet{mcquillanROTATIONPERIODS342014}. We use the same temperature calculation that we do for the larger TARS survey to allow a direct comparison against the default TARS catalog.
    \textit{Lower left:} The sample of \ktwo\ rotators from \citet{reinholdStellarRotationPeriods2020}.
    \textit{Lower right:} The sample of ZTF rotators from \citet{Lu2022}.
    }
    \label{fig:gyrochrone_plot}
\end{figure*}

\subsection{Additional types of variability}

The TARS catalog reveals significant structure in \teff\ and $M_G$ space, as we show in Figures~\ref{fig:rot_amp_cmd} and~\ref{fig:variability}. The Kraft break \citep{kraftStudiesStellarRotation1967} is apparent as a jump in the density of points and a change in the variability amplitude behavior around \teff$\gtrsim6600$\,K. 

The population of short-period variables above the Kraft break are unlikely to be from spots and stellar rotation. Instead, these are predominantly pulsators (e.g., $\delta$~Scuti and $\gamma$~Doradus) or other kinds of variables recovered by the Lomb-Scargle periodogram.

Populations of sub-giants and giants are also visible in the base catalog. These land well above the main-sequence with \teff\ generally between 4000\,K and 6000\,K. This population mostly vanishes in the high-quality catalog, however, most of the surviving targets land in or near the instability strip. Like those above the Kraft break, these are more likely to be pulsations or other sources of variability than spots and rotation.

To estimate the fraction of our measured periods that correspond to rotational modulation, we cross-matched the default TARS catalog with the \gaia\ DR3 \texttt{vari\_classifier\_result} table, which contains the variability classifications from \gaia\ DR3 \citep{2023A&A...674A..13E}. This cross-match returned \added{293{,}619} sources. Of these, \added{225{,}371} are classified as Solar-like modulation, \added{33{,}085} as RS CVn systems, \added{12{,}361} as young stellar objects (YSOs), \added{10{,}551} as eclipsing binaries, \added{8{,}995} as $\delta$ Scuti, $\gamma$ Doradus, or SX Phoenicis pulsators, and \added{2{,}465} as chemically peculiar Ap/Bp stars. The locations of these variability classes in rotation--temperature and color--magnitude space are shown in Figure~\ref{fig:variability}.

Assuming the \gaia\ DR3 classifications are correct and that the detected periods for Solar-like stars, RS CVn systems, YSOs, and Ap/Bp stars arise from rotational modulation, we find that approximately \added{93\%} of the matched sample corresponds to stellar rotation. We adopt this value as an empirical estimate of the fraction of rotation periods in our catalog.

\subsection{Unresolved Binaries}\label{sec:phot_binary_analysis}

Stars that appear brighter than the single-star main sequence on a color–magnitude diagram are likely to be spatially unresolved binaries.  This effect is evident in the top panels of Figure~\ref{fig:variability}, where a narrow overdensity above the main sequence is the binary sequence.

\citet{Simonian2019} quantified this effect in the \kepler\ rotator sample of \citet{mcquillanROTATIONPERIODS342014}, defining a photometric binary as a star lying more than 0.3 mag above the main sequence. They found that 59\% of fast rotators ($1.5 \leq P_{\rm rot} < 7$ days) met this criterion, compared to 29\% across the full sample, suggesting that most fast rotators are not young single stars, but are instead tidally synchronized binaries.

Here, we perform a similar analysis to estimate the photometric binary fraction in our sample. We begin by selecting a high-purity sample of rotators from the TARS catalog. We adopt stars with \added{$S_{\rm signal} > 0.8$ and $P(\rm match)$ or $P(\rm harmonic) > 0.8$}, enforce all quality flags \added{(no contaminants, no binaries, no multiple periods, period was measured in multiple TESS sectors)}, and remove \added{harmonic} periods rather than doubling them. These steps yield match rates exceeding 95\% across validation sets.

\begin{figure*}
    \centering
    \includegraphics[width=1\linewidth]{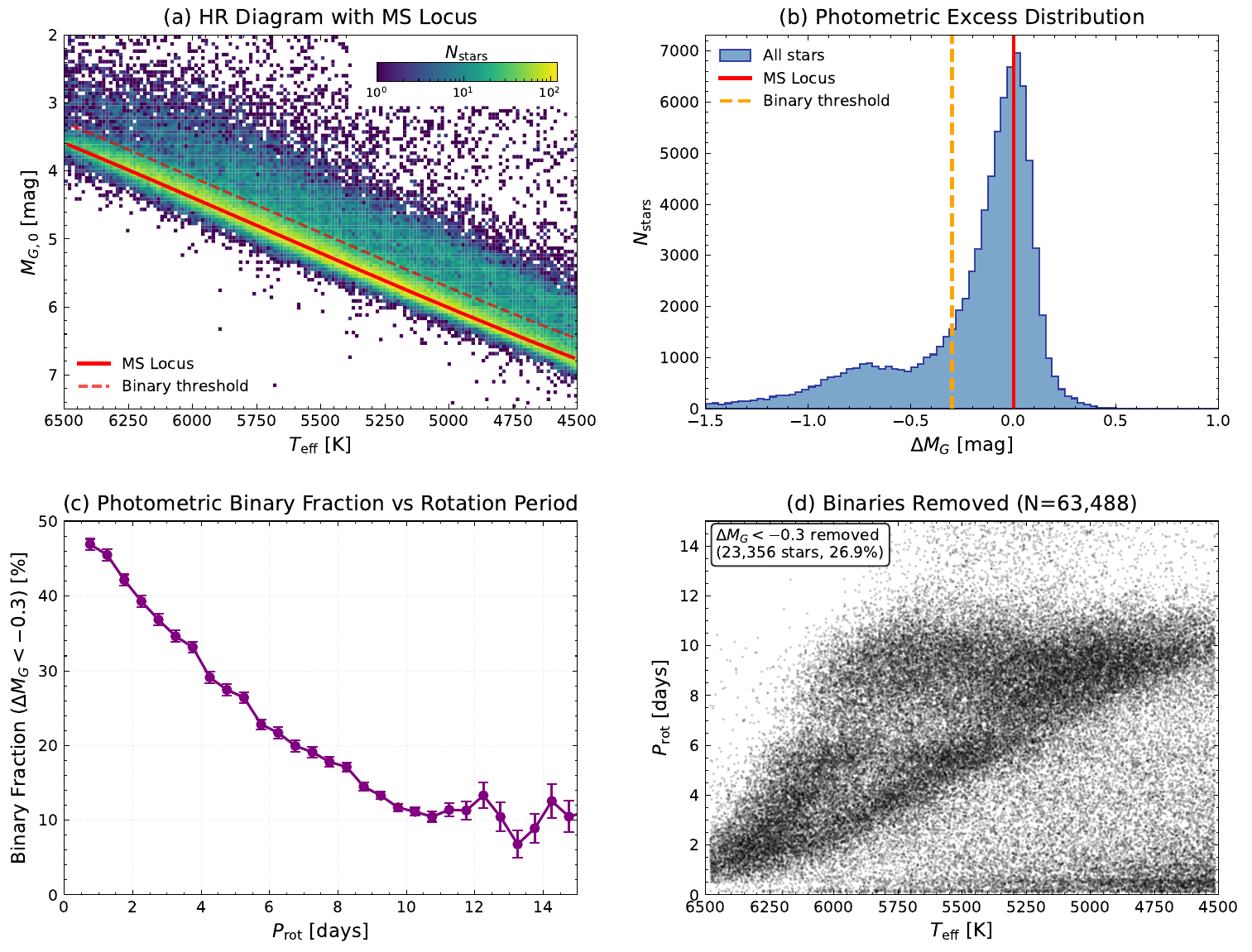}
    \caption{{\bf Photometric binarity is common at fast periods.}
    (a) HR diagram for stars with $4500 < T_{\rm eff} < 6500$ K, showing the empirical main-sequence locus (solid red line) and the $\Delta M_G = -0.3$ binary threshold (dashed red line). For this analysis, we have removed \added{harmonics} from the default TARS catalog instead of doubling them.
    (b) Distribution of photometric excess $\Delta M_G$ across the sample; the peak near $-0.75$ mag corresponds to equal-mass binaries.
    (c) Binary fraction as a function of rotation period, showing a strong excess among fast rotators.
    (d) Distribution of stars in $T_{\rm eff}$–$P_{\rm rot}$ space after binary removal.}
    \label{fig:phot_bin}
\end{figure*}

To define the single-star main sequence, we bin stars in $M_{G,0}$–$T_{\rm eff}$ space (using 60 bins), and compute the 70th percentile of $M_{G,0}$ in each bin. This choice prioritizes stars on the lower envelope of the CMD—fainter at fixed $T_{\rm eff}$ and thus more likely to be single. A spline fit to these percentile points defines the empirical main-sequence locus for $4500 < T_{\rm eff} < 6500$ K, avoiding cooler stars where the binary sequence blends with young single-star populations \citep{kerrStarsPhotometricallyYoung2021}. The result is shown in Figure~\ref{fig:phot_bin}(a) as the red curve.

We define the photometric excess as:
\begin{equation}
\Delta M_G = M_{G,0} - M_{G, \mathrm{locus}},
\end{equation}
where $M_{G,0}$ is the dereddened absolute magnitude and $M_{G, \mathrm{locus}}$ is the expected magnitude for a single star at that $T_{\rm eff}$.

The distribution of $\Delta M_G$ for stars in the cleaned sample with $4500 < T_{\rm eff} < 6500$ K is shown in panel (b) of Figure~\ref{fig:phot_bin}. The binary sequence peaks near $\Delta M_G \sim -0.75$, consistent with equal-mass binaries.

Adopting the threshold $\Delta M_G < -0.3$ to identify photometric binaries, we find that \added{26.9\%} of stars in our sample meet this criterion, comparable to the 28\% reported by \citet{Simonian2019}. Panel (c) shows that the binary fraction is strongly dependent on rotation period, exceeding 45\% among the fastest rotators and declining to $\sim5\%$ at longer periods. This provides additional evidence for the correlation between binarity and rapid rotation. Panel (d) shows the $T_{\rm eff}$–$P_{\rm rot}$ distribution after removing stars with $\Delta M_G < -0.3$.

\subsection{Structure in $P_{\rm rot}$--$T_{\rm eff}$ Diagram} \label{subsec:prot_teff_structure}

\begin{figure}
    \centering
    \includegraphics[width=1\linewidth]{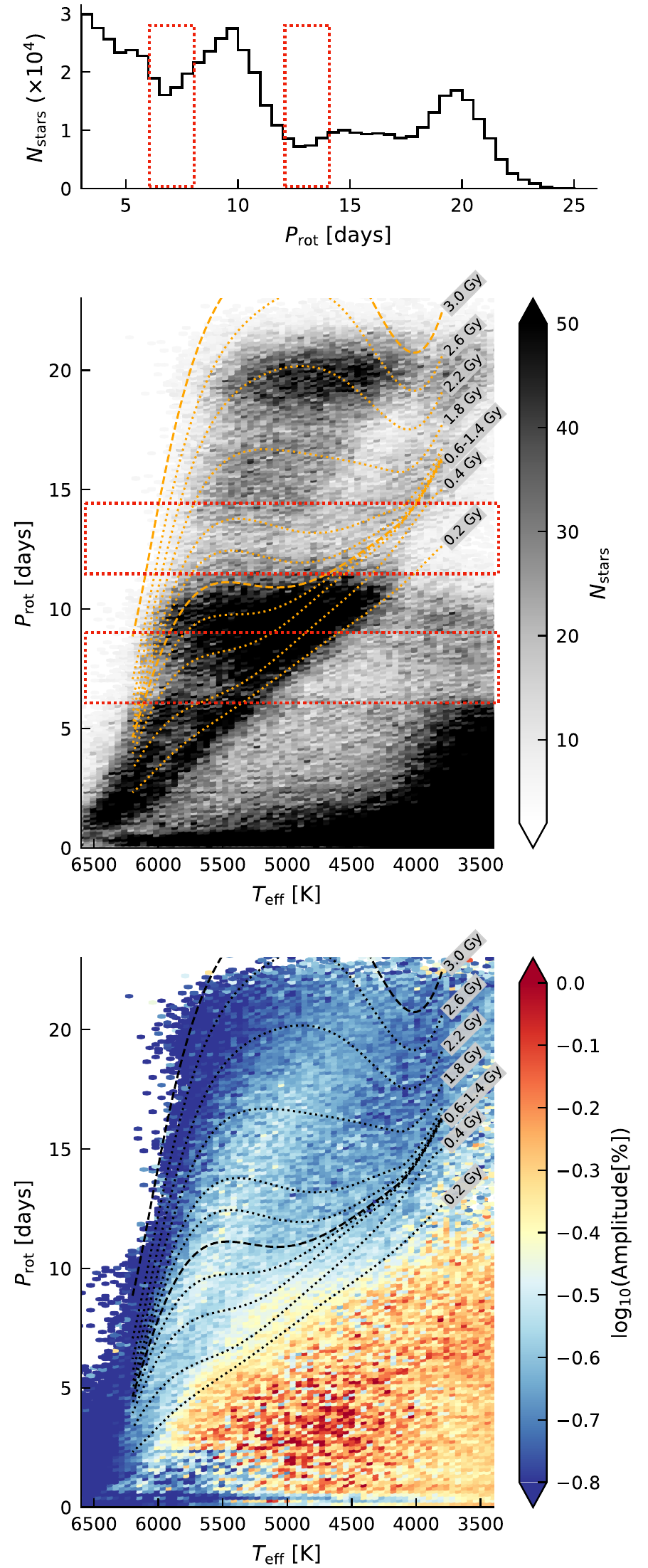}
    \caption{
    \textbf{The rotation-effective temperature distribution of the default TARS sample.}  Regions of interest are highlighted and discussed in Section~\ref{subsec:prot_teff_structure}.  Color bars are tuned to maximize contrast in the period gap.}
    \label{fig:prot_circled}
\end{figure}

The observed density of stars in the \prot--\teff\ plane encodes at least three separate effects: (1) the underlying stellar age distribution; (2) the mapping from age to rotation period and variability amplitude, including the effects of binarity; and (3) the survey's selection function, including both detection efficiency and biases in period recovery.

The middle panel of Figure~\ref{fig:prot_circled} shows the number density of TARS detections in the \prot--\teff\ plane.  The morphology is complex, with many over- and under-dense regions.  We discuss the most important regions below, and connect them to the median rotational amplitudes shown in the lower panel.

\subsubsection{Non-Astrophysical Effects Near Harmonics of TESS Orbital Period}

The top and middle panels of Figure~\ref{fig:prot_circled} show under-dense regions near $P_{\rm rot}\approx6.8$ and $13.6$~days, and over-dense regions near 10 and 20~days.  The under-dense regions align with the half lunar-month timescale of the TESS orbit and its first harmonic. 
We believe at least three effects contribute to these over- and under-dense regions.
First, in low-amplitude light curves, gap-related systematics can masquerade as periodic signals (Figure~\ref{fig:systematics}). Our systematics classifier rejects such signals, as most are not astrophysical.  However, the systematics classifier could be overly aggressive, due to the high density of real systematics.  Second, our \added{harmonic} classifier struggles to differentiate between matches and \added{harmonics} in the 6-8 day gap (see the top right panel of Figure~\ref{fig:rf_alias_thresholds}).   The incompleteness of the \added{harmonic} classifier in these regions contributes to the under-densities.
Finally, our measured periods near harmonics of the TESS orbital period suffer $\lesssim$5\% biases, revealed by the slight non-linearities in Figure~\ref{fig:kepler_k2_before_after}.  These tend to move stars from under- to over-dense regions, particularly near the 10~day and 20~day peaks. We note that the incidence of non-recovered periods from our validation samples is $<10\%$ through these overdensities, which suggests that most of the recovered periods in the overdensities are reliable.

\subsubsection{Astrophysical Period Gap Spanning F8V-M2V}

{\bf Cool Star Gap Recovered}: The ``intermediate period gap'' in the rotation period distribution for K and M dwarfs
has been documented in numerous studies of Kepler, K2, and ZTF data \citep[e.g.,][]{mcquillanROTATIONPERIODS342014,davenportRotatingStarsKepler2018,reinholdStellarRotationPeriods2020,gordonStellarRotationK22021,davidFurtherEvidenceModified2022,Lu2022}.
At $T_{\rm eff}\lesssim 5000\,$K, 
Figures~\ref{fig:gyrochrone_plot} and~\ref{fig:prot_circled} show that the TARS sample recovers this gap, and is most similar in stellar density to the Kepler and K2 samples.

Two possible explanations for the cool star gap have been previously discussed.  The first is that the gap coincides with a decrease in photometric variability amplitude \citep{Reinhold2019,reinholdStellarRotationPeriods2020}, which would reduce the detectability of rotation periods. 
In this first scenario, the physical origin of the amplitude decrease could be linked to the transition from spot- to faculae-dominated rotational modulation \citep[e.g.,][]{Shapiro2014}.
The alternative explanation invokes a phase of accelerated spin-down following core-envelope recoupling, during which stars rapidly evolve through intermediate rotation periods \citep{curtisTemporaryEpochStalled2019, gordonStellarRotationK22021, Lu2022}. The intermediate period gap identified in our sample coincides with a region of systematically lower variability amplitudes (Figure~\ref{fig:rot_amp_cmd}), consistent with both interpretations.

{\bf The Gap Extends to Sun-like Stars}:
The number density of K0V-F8V (5300-6200\,K) stars at $\lesssim$10\,days in Figure~\ref{fig:prot_circled} shows a diagonal underdensity: for TESS, these (roughly) Sun-like stars separate into a cool and a hot branch.
This underdensity appears to be an extension of the now-classic Kepler and K2 ``intermediate period gap'' to hotter stars.
The same region shows reduced photometric variability amplitudes in Figure~\ref{fig:rot_amp_cmd} and the lower panel of Figure~\ref{fig:prot_circled}, as a diagonal blue swath between two higher-amplitude branches, continuously spanning temperatures of 3800-6200~K.
While K2 previously showed local minima in variability amplitudes in 
hotter region \citep{reinholdStellarRotationPeriods2020}, 
the drop in apparent {\it number density} of stars is, to our knowledge, new to the TARS sample. 
Photometric binarity seems to not be related, since Figure~\ref{fig:phot_bin}d shows the same structure in a sample with photometric binaries removed.
Our \added{harmonic} handling is also not related, since the same figure includes no period doubling.

Why does TESS see this underdensity for Sun-like stars, when Kepler and K2 did not?  In \prot-\teff-amplitude space, the continuity of the diagonal low-amplitude gap implies a single physical cause, perhaps the spot- to faculae-dominated transition \citep{Reinhold2019}.  
In \prot-\teff-number density, there are two key distinguishing factors when comparing TESS and Kepler/K2.
The Kepler prime mission was biased against young stars since its average target was one scale height above the galactic plane, where kinematic heating lifts older stars \citep{Bouma2024gyro}.
TESS targets span all galactic latitudes, and so do not share this selection bias against young stars.
Separately, Kepler and K2 were also more sensitive to lower-amplitude photometric modulation, which would tend to fill in regions with low number density.
A quantitative reconstruction of the stellar age and number density distributions from TARS would be a worthy topic for future work.  However,
it is beyond our current scope.


\section{Discussion}\label{sec:discussion}

The TARS catalog is a search for periodic variability encompassing 7,481,412 stars observed in TESS Sectors 1--96 with $T < 16$ and distances $d < 500$ pc. With more than \added{1 million} period detections that pass our default selection, this catalog will be invaluable for a wide range of studies, including stellar spin-down, young associations, and stellar ages. 

\subsection{Effectiveness of Doubling Harmonics}

A major benefit of our catalog is the extension to longer-periods thanks to our \added{harmonic} classifier. With this, we are able to extend TESS rotation studies to stellar populations and age regimes that were previously inaccessible due to TESS’s short observational baseline. This improvement is illustrated by the rotation sequences of Praesepe and the Hyades shown in Figure~\ref{fig:praesepe_alias}. After applying the \added{harmonic} correction, the full rotation sequences of both clusters are recovered, enabling the construction of coherent rotation sequences for older clusters using TESS data alone. 

A similar effect can be seen in Figure~\ref{fig:gyrochrone_plot}, where the intermediate period gap for cool stars with long rotation periods ($P_{\rm rot} \gtrsim 13$~days) is also recovered. The presence and morphology of this gap closely match those seen in the \kepler, \ktwo, and ZTF samples, all of which benefit from longer observational baselines and are therefore less susceptible to the aliasing issues that affect TESS. 

The downside of the classifier is that it sometimes incorrectly doubles the period. We have made the decision to use the \added{harmonic} classifier to double half-period \added{harmonics} (which can comprise up to $\sim30\%$ of detections). This virtually eliminates half-period \added{harmonics} but comes at the cost of $\sim3\%$ of the sample being erroneously doubled.  Alternative options for an interested user include using the code on Zenodo to remove \added{harmonics} instead of doubling them, or manually inspecting the vetting plots (Figure~\ref{fig:first_stage_lcs}).

\subsection{Areas For Caution}

Several caveats should be kept in mind when interpreting the TARS catalog and the structure of the \prot--\teff\ plane. First, blending and residual contamination from neighboring stars can introduce spurious periods (Section~\ref{subsec:quality_flags}), even when the star is multiple TESS pixels away from the target star \citep[cf.][Figure~7]{Stafford2026}. Users interested in rotation periods for stars in crowded regions such as the galactic plane should take care to account for potential contamination.

Second, our \added{harmonic}-doubling strategy substantially reduces half-period contamination but introduces a small ($\sim$2--3\%) residual population of incorrectly doubled periods (Section~\ref{subsubsec:alias_rf}). In addition, mild period-recovery biases and harmonics of the TESS window function imprint under- and over-densities near $\sim$6–8, $\sim$10, and $\sim$20 days (Section~\ref{subsec:prot_teff_structure}). While our validation samples show that the periods in these regions are reliable, users could take the additional step of manually vetting individual periods of interest, including by adopting their own criteria for treating multi-sector information.

\subsection{Future Directions}

\textit{Longer rotation periods.} 
The present analysis derives rotation periods from individual TESS sectors. Recent studies \citep{Claytor_2024, Hattori2025} have demonstrated that rotation periods up to $\sim80$ days can be recovered in the TESS continuous viewing zones by combining multi-sector data. Extending TARS to incorporate multi-sector light curves would therefore enable the recovery of longer periods. A particularly promising avenue would be to adapt the methodology of \citet{Hattori2025} to non-consecutive sectors, testing whether coherent long-period signals can be reconstructed even when observations are temporally separated. Because most TARS targets were observed in multiple sectors, such an extension could enable long-period measurements for a significant fraction of the sample.

\textit{Improved classification.} 
Our \added{harmonic} classifier currently misidentifies $\sim15\%$ of rotation periods as doubled periods without adopting a stricter $P(\rm harmonic)$ or $P(\rm match)$ threshold, like we do in constructing our default catalog. This behavior suggests partial overlap between true matches and \added{harmonics} in feature space, limiting separability. Developing additional or more discriminating features --- particularly those sensitive to harmonic structure or phase coherence --- could improve class separation and reduce systematic doubling errors. Refining the classifier would increase both completeness and reliability, especially for moderate-amplitude signals.

\textit{Variability classification.} 
Although we estimate that \added{$\sim93\%$} of reported periods correspond to stellar rotation, we have not attempted to subdivide the catalog by variability type. Implementing a dedicated variability classifier (for example, an additional random forest trained to distinguish rotational modulation from pulsation, eclipses, or other variability) would produce a cleaner rotation-only subset. Such a classification would both enhance rotation-focused studies and facilitate investigations of non-rotational variability within the broader TARS sample.

\section{Conclusion}\label{sec:conclusion}

We have conducted an all-sky rotation survey with TESS by searching 7,481,412 unique stars for stellar variability across 39,061,674 TESS light curves. The main takeaways are as follows.

\begin{enumerate}[leftmargin=12pt,topsep=0pt,itemsep=0ex,partopsep=1ex,parsep=1ex]
    \item We have produced a catalog of \tarsnrots\ measured periods, $\approx$93\% of which are from stellar rotation.  This expands the number of stars with known rotation periods by a factor of \added{2.3} within 100~pc and \added{4.0} within 500~pc.  The most common types of other variability include eclipsing binaries, $\delta$~Scuti pulsators, and $\gamma$~Doradus pulsators.
    \item Our methodology allows us to remove half-period \added{harmonics} and recover reliable rotation periods as long as 25~days from a single TESS sector.
    \item We provide a quantitative framework for evaluating the completeness and reliability of this catalog and show that our periods largely agree with rotation periods from \kepler, \ktwo, and ZTF.
    \item We recover the intermediate period gap for stars cooler than 5000\,K with \prot$\ge$13 days, matching the morphology seen in Kepler, K2, and ZTF. The detection of this gap in TESS data, facilitated by our harmonic-correction framework, demonstrates that TARS can be used to study the angular momentum evolution of stars well into their intermediate-age ``stalled'' spin-down phase.
    \item We observe an extension of the intermediate period gap to hotter stars, which manifests as two separate \prot-\teff\ branches in the number density of K0V-F8V (5300-6200\,K) stars at $\lesssim$10\,days.  Both the hot and cool regions of the period gap overlap with a local minimum in rotation amplitudes, and so probably share a common origin.
    \item We confirm a distinct transition in rotational behavior and variability amplitude near the Kraft break at \teff$\simeq6600$\,K. This provides an empirical benchmark for the disappearance of deep convective envelopes in massive stars.
    \item By analyzing the color-magnitude distribution of our rotators, we confirm that  a significant portion of rapid field rotators ($<7$\,d) are tidally synchronized binaries. The binary fraction among the fastest rotators exceeds 45\%, compared to $\simeq5\%$ for stars on the slow-rotator sequence.
\end{enumerate}

We recognize that users may have different science needs and that the default catalog we present may not be sufficient for each science goal. To that end, we have written a command line utility that allows a user to rebuild our catalog based on the different systematics classifier, \added{harmonic} classifier, and quality flagging thresholds appropriate for their science. This script will also estimate completeness and reliability statistics for the user-defined inputs based on the same validation samples in Section~\ref{sec:validation}. This script is available on Zenodo\footnote{\href{https://doi.org/10.5281/zenodo.19917941}{10.5281/zenodo.19917941}}. Our light curves and vetting plots are available at MAST as a High Level Science Product\footnote{\url{https://archive.stsci.edu/hlsp/tars}} via \dataset[10.17909/2sff-fn29]{\doi{10.17909/2sff-fn29}}. Each of our 39 million sector level measurements are also available on Zenodo and follow the table format described in Appendix~\ref{appendix:full_table}.

\begin{acknowledgments}

\added{The authors would like to thank Mara Bernizzoni for helpful comments on this manuscript}. This material is based upon work supported by the National Science Foundation Graduate Research Fellowship Program under Grant No. DGE-2439854. Any opinions, findings, and conclusions or recommendations expressed in this material are those of the authors and do not necessarily reflect the views of the National Science Foundation. AWB thanks the LSST-DA Data Science Fellowship Program, which is funded by LSST-DA, the Brinson Foundation, the WoodNext Foundation, and the Research Corporation for Science Advancement Foundation; his participation in the program has benefited this work. 

LGB gratefully acknowledges support from the Carnegie Fellowship. 

AWM and AWB were funded by a grant from NASA's Astrophysics Data Analysis program (ADAP 80NSSC24K0619). AWM was further supported by a grant from the NSF CAREER program (AST-2143763).

This research was done using services provided by the OSG Consortium \citep{osg07, osg09, https://doi.org/10.21231/906p-4d78, https://doi.org/10.21231/0kvz-ve57}, which is supported by the National Science Foundation awards \#2030508 and \#2323298. 

This paper includes data collected by the TESS mission. Funding for the TESS mission is provided by the NASA's Science Mission Directorate. The TESS data used in this paper can be found at MAST \citep{https://doi.org/10.17909/0cp4-2j79}.

This work has made use of data from the European Space Agency (ESA) mission
{\it Gaia} (\url{https://www.cosmos.esa.int/gaia}), processed by the {\it Gaia}
Data Processing and Analysis Consortium (DPAC,
\url{https://www.cosmos.esa.int/web/gaia/dpac/consortium}). Funding for the DPAC
has been provided by national institutions, in particular the institutions
participating in the {\it Gaia} Multilateral Agreement.

\end{acknowledgments}

\begin{contribution}

Per https://credit.niso.org: Conceptualization: All authors. Data curation: AWB. Formal analysis: All authors. Funding acquisition: All authors. Investigation: AWB. Methodology: All authors. Project administration: All authors. Resources: All authors. Software: AWB, LGB. Supervision: AWM, LGB. Validation: AWB, LGB. Visualization: AWB, LGB. Writing – original draft: AWB. Writing – review \& editing: all authors.

\end{contribution}

\facilities{TESS \citep{Ricker2015}, Gaia \citep{2016A&A...595A...1G, 2023A&A...674A...1G}, Kepler \citep{Borucki2010}, K2 \citep{Howell2014}, ZTF \citep{Bellm2019}}

\software{astropy (\!\citealt{astropy:2013, astropy:2018, astropy:2022}),  
tess-point \citep{2020ascl.soft03001B},
matplotlib \citep{hunter2007matplotlib},
pandas \citep{mckinney-proc-scipy-2010, reback2020pandas},
unpopular \citep{hattoriUnpopularPackageDatadriven2021},
scikit-learn \citep{scikit-learn}
          }

\appendix

\section{Lomb-Scargle Implementation and Calculated Parameters} \label{sec:measurements}

We computed the Lomb Scargle periodogram using a linearly spaced grid in frequency from 0.1 to 20 days with one million points. This is over-sampled relative to the available Nyquist information content \citep{vanderplasUnderstandingLombScargle2018a} but keeps gaps between grid points below the measurement uncertainties, even at the longest periods. 

We pre-process our data by performing sigma clipping to remove points more than $3\sigma$ away from the median light curve value. This step removes outliers that could otherwise alter the results.

To model the dominant periodic signal, we fit both single- and two-term Lomb-Scargle models. The one-term Lomb-Scargle model captures simple sinusoidal variability and yields the amplitude, phase, and power of the best-fit sinusoid. However, stellar rotation signals often deviate from purely sinusoidal shapes. In such cases, a two-term model, which includes the first harmonic, can better describe non-sinusoidal light curve morphologies, such as double-dip modulations. The two term model is not used for our periodogram powers.

The full list of our recorded parameters is as follows:

\begin{itemize}

\item \texttt{first\_period}, \texttt{second\_period}, \texttt{third\_period}, \texttt{fourth\_period}, \texttt{fifth\_period}:
Periods (in days) of the top five Lomb--Scargle (LS) periodogram peaks after restricting to periods
$P < 20\,\mathrm{d}$ and enforcing each period to be $>10\%$ away from a previously recorded period. 
Periods greater than 20 days are only used for calculating the power at twice the period of the highest peak in the periodogram (see \texttt{power\_twice\_period} below).

\item \texttt{first\_power}, \texttt{second\_power}, \texttt{third\_power}, \texttt{fourth\_power}, \texttt{fifth\_power}:
LS powers evaluated at the corresponding peak periods. The periodogram is computed on sigma-clipped fluxes and using a scalar flux uncertainty estimated from the point-to-point RMS (see \texttt{p2p\_rms}).

\item \texttt{first\_snr}, \texttt{second\_snr}, \texttt{third\_snr}, \texttt{fourth\_snr}, \texttt{fifth\_snr}:
A simple periodogram signal-to-noise-like metric for each peak, defined as the ratio of the peak power to the median power across the
(full) computed periodogram: $\mathrm{SNR} \equiv \frac{\mathrm{power}}{\mathrm{median}(\mathrm{power})}$.

\item \texttt{best\_fit\_amplitude}:
Amplitude of the best-fitting single-term LS model at the best frequency $f_{\rm best}=1/P_{\rm best}$.
If the LS model parameters at $f_{\rm best}$ are $\boldsymbol{\theta}$ with sine and cosine coefficients
$\theta_1$ and $\theta_2$, the reported amplitude is $A \equiv \sqrt{\theta_1^2+\theta_2^2}$.

\item \texttt{time\_first\_minimum}:
Time of the first local minimum of the best-fitting LS model light curve, i.e., the earliest $t$ at which the model
$y_{\rm fit}(t)$ has a local minimum. This is obtained by finding troughs in $y_{\rm fit}(t)$ and returning the
time of the first such trough.

\item \texttt{first\_peak\_width}:
Width (in days) of the dominant LS peak in \emph{period} space, measured at half-maximum using
\texttt{scipy.signal.peak\_widths}. Concretely, the width is computed in frequency space at half-height,
converted to periods, and differenced: $w_P \equiv P_{\rm right}-P_{\rm left}$.

\item \texttt{p2p\_rms}:
Point-to-point RMS noise proxy computed from successive flux differences
$\Delta F_j \equiv F_{j}-F_{j+1}$.
Let $\widetilde{\Delta F}\equiv \mathrm{median}(\Delta F)$ and define centered differences
$\Delta F'_j=\Delta F_j-\widetilde{\Delta F}$.
The reported value averages the upper and lower robust dispersions: $\mathrm{p2p} \equiv \frac{1}{2}\left[(Q_{84}-Q_{50}) + (Q_{50}-Q_{16})\right]$, where $Q_p$ denotes the $p$th percentile of the sorted $\Delta F'_j$ distribution.

\item \texttt{a\_90\_10}:
Flux amplitude defined as the 90th--10th percentile range of the fluxes: $A_{90-10} \equiv Q_{90}(F)-Q_{10}(F).$

\item \texttt{red\_chi2}:
Reduced chi-square of the best-fitting single-term LS model at $f_{\rm best}$, using the point-to-point based scatter as the
uncertainty scale. With residuals $r_i\equiv F_i-y_{\rm fit}(t_i)$,
\[
\chi^2 \equiv \sum_{i=1}^{N}\left(\frac{r_i}{\mathrm{p2p\_rms}(F)}\right)^2,\qquad
\chi^2_\nu \equiv \frac{\chi^2}{N-3},
\]
where $N$ is the number of time points and $N-3$ is the assumed number of degrees of freedom.

\item \texttt{snr}:
A heuristic signal-to-noise metric combining relative flux amplitude, noise, time baseline, and the inferred period:
\[
\mathrm{snr} \equiv \left(\frac{A_{90-10}}{\mathrm{p2p}}\right)\sqrt{\frac{T_{\rm base}}{P_{\rm best}}},
\]
where $T_{\rm base}\equiv \max(t)-\min(t)$ is the time baseline and $P_{\rm best}=\texttt{first\_period}$.

\item \texttt{power\_twice\_period}:
LS power evaluated at the frequency corresponding to twice the best period, i.e.\ at
$f = 0.5f_{\rm best}$ (equivalently, $P=2P_{\rm best}$). The code finds the
closest sampled frequency to $0.5/P_{\rm best}$ and records the power at that index.

\item \texttt{one\_term\_bic}:
Bayesian Information Criterion (BIC) for the single-term sinusoidal model fit at $f_{\rm best}$, computed as
\[
\mathrm{BIC}_1 \equiv \chi^2_1 + k_1\ln N,\qquad k_1=3,
\]
with
\[
\chi^2_1 \equiv \sum_{i=1}^{N}\frac{\left(F_i-y_{\rm fit}(t_i)\right)^2}{\sigma_F^2},
\]
where $\sigma_F\equiv \mathrm{p2p\_rms}(F)$ is treated as a per-point flux uncertainty.

\item \texttt{two\_term\_bic}:
BIC for a two-term ($n_{\rm terms}=2$) LS model evaluated at half the best frequency ($0.5f_{\rm best}$): $\mathrm{BIC}_2 \equiv \chi^2_2 + k_2\ln N$, where $\chi^2_2$ is computed analogously using the two-term model prediction, and $k_2=5$ is the parameter penalty adopted in the code for $n_{\rm terms}=2$.

\item \texttt{two\_term\_bic\_same\_period}:
Same as \texttt{two\_term\_bic}, but evaluating the two-term model at $f_{\rm best}$ (i.e.\ enforcing the same
period as the primary peak).

\item \texttt{delta\_bic}:
Difference in BIC between the two-term (at $0.5 f_{\rm best}$) and one-term model: $\Delta \mathrm{BIC} \equiv \mathrm{BIC}_2 - \mathrm{BIC}_1.$

\end{itemize}
The following metrics are used only in the \added{harmonic} classifier:

\begin{itemize}
\item \texttt{period\_ratio\_1\_2}:
The ratio of \texttt{first\_period} divided by \texttt{second\_period}.
\item \texttt{period\_ratio\_1\_3}:
The ratio of \texttt{first\_period} divided by \texttt{third\_period}.
\item \texttt{snr\_ratio\_1\_2}:
The ratio of \texttt{first\_snr} over \texttt{second\_snr}.
\item \texttt{power\_ratio\_1\_2}:
The ratio of \texttt{first\_power} over \texttt{second\_power}.

\end{itemize}

\section{Random Forest Validation Plots} \label{sec:rf_val_plots}

\begin{figure*}[!t]
    \centering
    \includegraphics[width=0.9\linewidth]{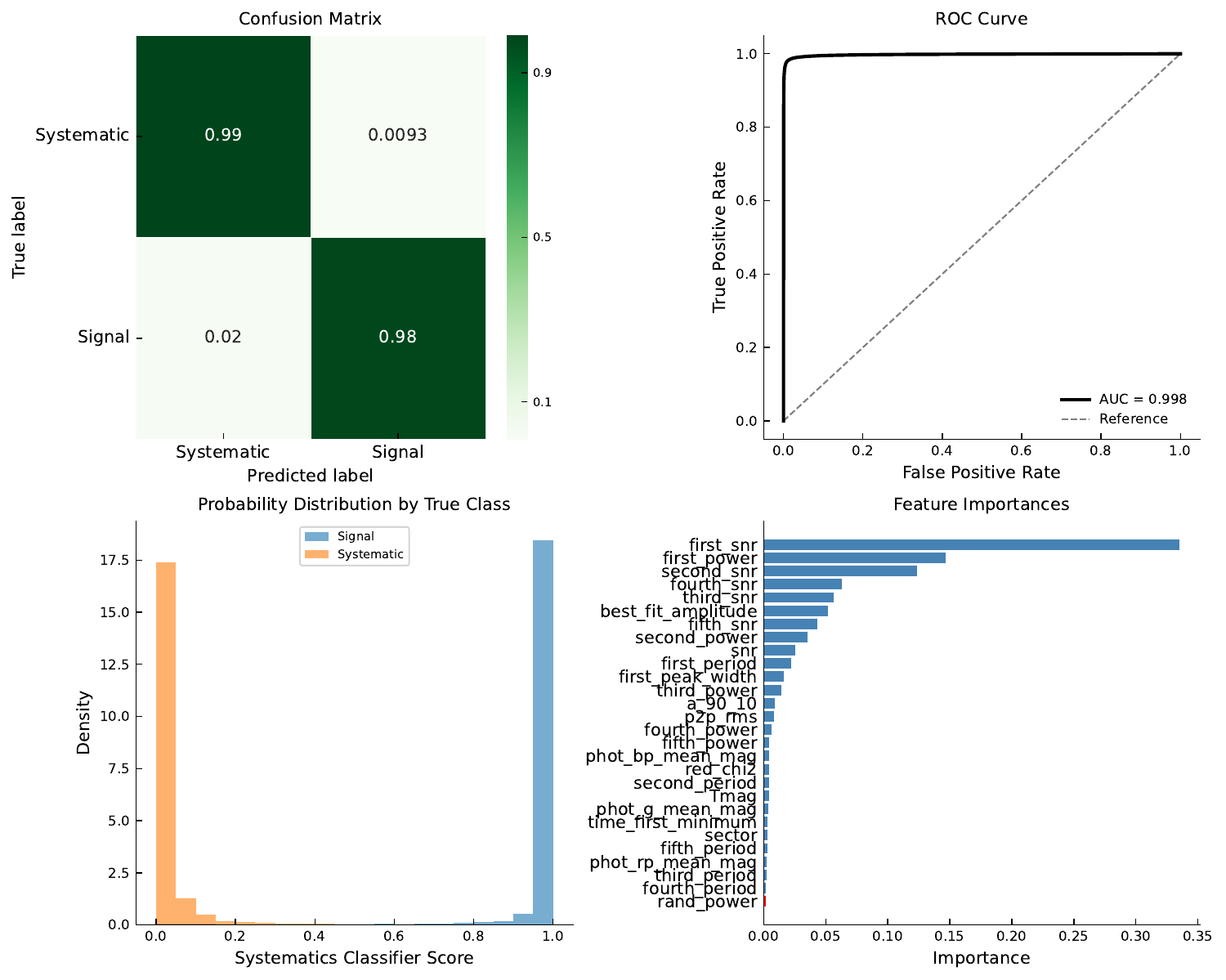}
    \caption{Extra validation plots for our systematics classifier.
    \textit{Top left:} Confusion matrix for our negative (systematic) and positive (rotation signal) classes. 
    \textit{Top right:} ROC curve for this classifier showing the effectiveness of the classifier at distinguishing systematics from probable rotation signals.
    \textit{Bottom left:} Histogram showing the probability distribution of each class. The bimodal distribution with a wide valley between peaks indicates strong classifier performance.
    \textit{Bottom right:} The feature importances from our systematics classifier, indicating that the signal-to-noise ratio of the first peak in the periodogram, the power of the first peak in the periodogram, and SNR of the second peak in the periodogram, the amplitude of the light curve, and the SNR of the third peak in the periodogram are most effective for disentangling systematics and true rotation signals. We additionally plot in red a randomized variable (\texttt{rand\_power}), which is made by taking the \texttt{first\_power} values and randomly shuffling them. \texttt{rand\_power} has the lowest feature importance of all input features.
    }
    \label{fig:systematics_validation_plots}
\end{figure*}

\begin{figure*}[!t]
    \centering
    \includegraphics[width=0.9\linewidth]{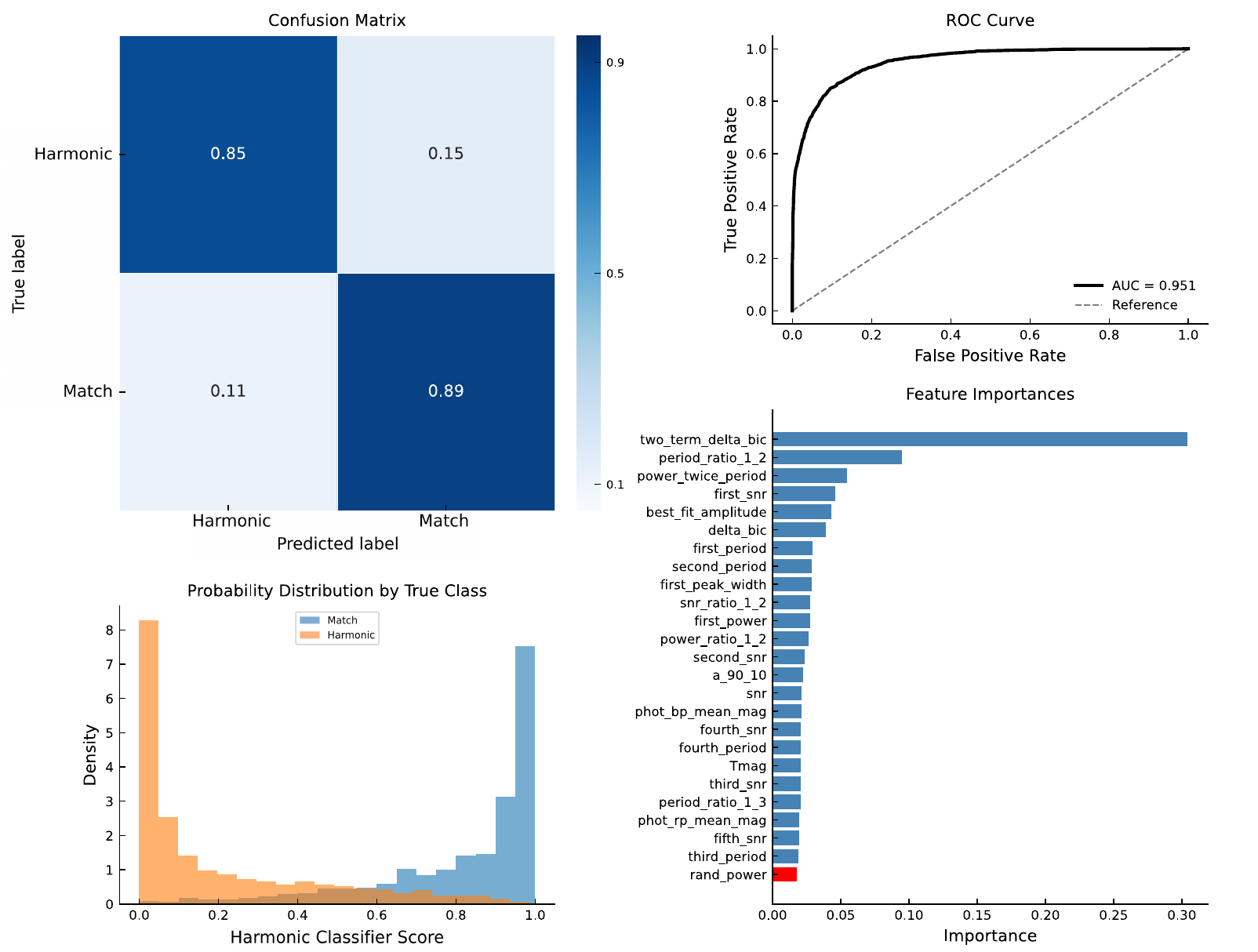}
    \caption{Extra validation plots for our \added{harmonic} classifier.
    \textit{Top left:} Confusion matrix for our negative (\added{harmonic}) and positive (true rotation signal) classes. 
    \textit{Top right:} ROC curve for this classifier showing the effectiveness of the classifier at distinguishing \added{harmonics} from \added{non-harmonic} rotation signals.
    \textit{Bottom left:} Histogram showing the probability distribution of each class. Compared to the systematics classifier, there is more overlap in the valley between the bimodality, thought high probability cuts still ensure a clean sample.
    \textit{Bottom right:} The feature importances from our classifier, indicating that the difference in fit between a two-term sinusoidal model evaluated at the period from the highest peak in the periodogram and the a fit evaluated at twice that period carries high diagnostic power. The same randomized power variable as in Figure~\ref{fig:systematics_validation_plots} is plotted in red.
    }
    \label{fig:alias_validation_plots}
\end{figure*}

In this section we provide additional validation plots for our systematics classifier (Section~\ref{subsubsec:systematics}, Figure~\ref{fig:systematics_validation_plots}) and \added{harmonic} classifier (Section~\ref{subsubsec:alias_rf}, Figure~\ref{fig:alias_validation_plots}). For additional details on how each classifier was trained, please refer back to Section~\ref{subsec:validating_prot}.

\newpage
\section{Full Catalog Column Descriptions}\label{appendix:full_table}

\begin{longtable}{ccc}
    \caption{Periodogram results for 39 million single-sector TARS light curves.  Most entries do not correspond to detections of astrophysical variability.  A unique star has one row per observed TESS sector. } \label{tab:sector_results} \\
    \hline \hline
    Parameter & Example Value & \textbf{Description} \\
    \hline
    \endfirsthead

    \multicolumn{3}{c}{\tablename\ \thetable{} -- continued from previous page} \\
    \hline \hline
    Parameter & Example Value & \textbf{Description} \\
    \hline
    \endhead

    \hline
    \multicolumn{3}{r}{Continued on next page} \\
    \endfoot

    \hline
    \multicolumn{3}{l}{\parbox{0.95\textwidth}{\footnotesize \textbf{Note.} This table is published in its entirety in machine-readable format. One entry is shown for guidance regarding form and content.}} \\
    \endlastfoot

    \texttt{TICID} & 92634310 & TESS Input Catalog identifier \\
    \texttt{sector} & 82 & TESS sector number \\
    \texttt{first\_period} & 2.639 & Period of highest periodogram peak (days) \\
    \texttt{first\_power} & 0.03767 & Lomb-Scargle power of highest peak \\
    \texttt{second\_period} & 3.108 & Period of second-highest peak (days) \\
    \texttt{second\_power} & 0.00711 & Lomb-Scargle power of second-highest peak \\
    \texttt{third\_period} & 2.291 & Period of third-highest peak (days) \\
    \texttt{third\_power} & 0.00206 & Lomb-Scargle power of third-highest peak \\
    \texttt{fourth\_period} & 7.449 & Period of fourth-highest peak (days) \\
    \texttt{fourth\_power} & 0.00170 & Lomb-Scargle power of fourth-highest peak \\
    \texttt{fifth\_period} & 0.109 & Period of fifth-highest peak (days) \\
    \texttt{fifth\_power} & 0.00156 & Lomb-Scargle power of fifth-highest peak \\
    \texttt{first\_snr} & 286.95 & Signal-to-noise ratio of highest peak (peak power / median power) \\
    \texttt{second\_snr} & 54.15 & Signal-to-noise ratio of second-highest peak \\
    \texttt{third\_snr} & 15.72 & Signal-to-noise ratio of third-highest peak \\
    \texttt{fourth\_snr} & 12.94 & Signal-to-noise ratio of fourth-highest peak \\
    \texttt{fifth\_snr} & 11.87 & Signal-to-noise ratio of fifth-highest peak \\
    \texttt{best\_fit\_amplitude} & 0.00316 & Amplitude of best-fit sinusoid at \texttt{first\_period} (relative flux) \\
    \texttt{flag\_periodogram\_binary} & False & Flagged as possible binary from periodogram \\
    \texttt{time\_first\_minimum} & 3533.44 & Time of first minimum of best-fit sinusoid (days) \\
    \texttt{first\_peak\_width} & 0.238 & Full width at half maximum of highest peak (days) \\
    \texttt{p2p\_rms} & 0.01546 & Point-to-point RMS of light curve (relative flux) \\
    \texttt{a\_90\_10} & 0.02848 & 90th minus 10th percentile of flux distribution (relative flux) \\
    \texttt{red\_chi2} & 0.514 & Reduced $\chi^2$ of best-fit sinusoid \\
    \texttt{snr} & 5.765 & Signal-to-noise ratio of light curve at \texttt{first\_period} \\
    \texttt{power\_twice\_period} & 0.00006 & Periodogram power at twice \texttt{first\_period} \\
    \texttt{one\_term\_bic} & 5611.85 & BIC for single-term sinusoid model at \texttt{first\_period} \\
    \texttt{two\_term\_bic} & 5628.62 & BIC for two-term sinusoid model at $2\times$\texttt{first\_period} \\
    \texttt{delta\_bic} & 16.78 & $\Delta$BIC between two-term and one-term models \\
    \texttt{two\_term\_bic\_same\_period} & 5630.40 & BIC for two-term sinusoid model at \texttt{first\_period} \\
    \texttt{flag\_bad\_coverage} & False & Flagged for insufficient temporal coverage \\
    \texttt{dr2\_source\_id} & 2053956410104896640 & Gaia DR2 source identifier \\
    \texttt{dr3\_source\_id} & 2053956410104896640 & Gaia DR3 source identifier \\
    \texttt{ra} & 303.054 & Gaia DR3 right ascension (deg) \\
    \texttt{dec} & 31.288 & Gaia DR3 declination (deg) \\
    \texttt{pmra} & 29.467 & Gaia DR3 proper motion in right ascension (mas\,yr$^{-1}$) \\
    \texttt{pmdec} & $-72.841$ & Gaia DR3 proper motion in declination (mas\,yr$^{-1}$) \\
    \texttt{parallax} & 3.363 & Gaia DR3 parallax (mas) \\
    \texttt{radial\_velocity} & $-74.991$ & Gaia DR3 radial velocity (km\,s$^{-1}$) \\
    \texttt{l} & 69.582 & Galactic longitude (deg) \\
    \texttt{b} & $-1.475$ & Galactic latitude (deg) \\
    \texttt{phot\_g\_mean\_mag} & 14.052 & Gaia $G$-band apparent magnitude (mag) \\
    \texttt{phot\_rp\_mean\_mag} & 13.283 & Gaia $G_{\rm RP}$-band apparent magnitude (mag) \\
    \texttt{phot\_bp\_mean\_mag} & 14.686 & Gaia $G_{\rm BP}$-band apparent magnitude (mag) \\
    \texttt{ruwe} & 1.130 & Gaia Renormalized Unit Weight Error \\
    \texttt{distance} & 297.318 & Distance (pc) \\
    \texttt{extinction\_a0} & 0.105 & Monochromatic extinction at 550\,nm (mag) \\
    \texttt{extinction\_a0\_err} & 0.005 & Uncertainty on monochromatic extinction (mag) \\
    \texttt{phot\_g\_mean\_mag\_0} & 6.602 & Absolute $G$-band magnitude, corrected for extinction (mag) \\
    \texttt{phot\_rp\_mean\_mag\_0} & 5.851 & Absolute $G_{\rm RP}$ magnitude, corrected for extinction (mag) \\
    \texttt{phot\_bp\_mean\_mag\_0} & 7.213 & Absolute $G_{\rm BP}$ magnitude, corrected for extinction (mag) \\
    \texttt{BpmRp0} & 1.362 & Extinction-corrected $G_{\rm BP} - G_{\rm RP}$ color (mag) \\
    \texttt{teff} & 4511.8 & Calculated effective temperature (K) \\
    \texttt{non\_single\_star} & 0 & Gaia non-single star flag (0 = single, $>$0 = non-single) \\
    \texttt{camera} & 2 & TESS camera number (1--4) \\
    \texttt{ccd} & 2 & TESS CCD number (1--4) \\
    \texttt{Tmag} & 13.358 & TESS magnitude (mag) \\
    \texttt{num\_contams} & 0 & Number of contaminating sources within aperture \\
    \texttt{period\_ratio\_1\_2} & 0.849 & Ratio of \texttt{first\_period} to \texttt{second\_period} \\
    \texttt{period\_ratio\_1\_3} & 1.152 & Ratio of \texttt{first\_period} to \texttt{third\_period} \\
    \texttt{power\_ratio\_1\_2} & 5.299 & Ratio of \texttt{first\_power} to \texttt{second\_power} \\
    \texttt{snr\_ratio\_1\_2} & 5.299 & Ratio of \texttt{first\_snr} to \texttt{second\_snr} \\
    \texttt{two\_term\_delta\_bic} & $-1.773$ & $\Delta$BIC between two-term models at $2\times$ and $1\times$ \texttt{first\_period} \\
    \texttt{contamratio} & 0.137 & TICv8.2 contamination ratio \\
    \texttt{systematic\_score} & 0.00 & Systematics score \\
    \texttt{signal\_score} & 1.00 & Signal score \\
    \texttt{match\_score} & 0.83 & Match score \\
    \texttt{harmonic\_score} & 0.17 & Harmonic score \\
    \texttt{match\_prob} & 0.99 & Probability that the measured period is a match \\
    \texttt{harmonic\_prob} & 0.01 & Probability that the measured period is a half-period harmonic \\
    \texttt{n\_secs\_available} & 9 & Number of sectors in which star was observed (sectors 1–96 only) \\
\end{longtable}

\clearpage

\bibliography{PAPER-Prot_Bayes, Mannbib}{}
\bibliographystyle{aasjournalv7}

\end{document}